\documentclass[twocolumn]{revtex4}
\usepackage{type1cm}
\usepackage[dvipdfm]{graphicx}

\usepackage{amsmath,amssymb,bm}

\def\beq{\begin{equation}}
\def\eeq{\end{equation}}
\def\<{\langle}
\def\>{\rangle}
\def\rt#1{\sqrt{\mathstrut #1}}
\def\Tr{{\rm Tr}}

\renewcommand{\d}{\partial}

\begin{document}
\title{Instability of the mean-field states and generalization of phase separation in long-range interacting systems}
\author{Takashi Mori}
\email{mori@spin.phys.s.u-tokyo.ac.jp}
\affiliation{
Department of Physics, Graduate School of Science,
The University of Tokyo, Bunkyo-ku, Tokyo 113-0033, Japan
}

\begin{abstract}
Equilibrium properties of long-range interacting systems on lattices are investigated.
There was a conjecture by Cannas {\it et al.} [Phys. Rev. B {\bf 61}, 11521 (2000)] 
that the mean-field theory is exact for spin systems with nonadditive long-range interactions.
This is called ``exactness of the mean-field theory''.
We show that the exactness of the mean-field theory holds for systems on a lattice with non-additive two body long-range interactions
in the canonical ensemble with unfixed order parameters.
We also show that in a canonical ensemble with fixed order parameters,
exactness of the mean-field theory does not hold in one parameter region, which we call the ``non-mean-field region.''
In the non-mean-field region, an inhomogeneous configuration appears,
in contrast to the uniform configuration in the region where the mean-field theory holds.
This inhomogeneous configuration is not the one given by the standard phase separation.
Therefore, the mean-field picture is not adequate to describe these states.
We discuss phase transitions between the mean-field region and the non-mean-field region.
Exactness of the mean-field theory in spin glasses is also discussed.
\end{abstract}
\maketitle


\section{Introduction}

Long-range interactions cause several peculiar features:
negative specific heat in the microcanonical ensemble~\cite{Thirring1970}, long-lived metastable states~\cite{Griffiths1966},
and ensemble inequivalence~\cite{Barre2001}.
The nonadditivity of the interaction potential causes such anomalous properties.
When the integral of an interaction potential diverges at the large distances (i.e., 
a pair interaction behaves like $1/r^{\alpha}$ with $\alpha<d$ where $d$ is the dimension of space) or
when the interaction range is of the order of the system size,
the system cannot be divided into thermodynamically independent subsystems.
In this case, the system is said to be nonadditive.

Until now, the statistical physics of long-range interacting systems have attracted attention~\cite{Lecture_notes2002,Les_Houches_long}.
To understand the statistical mechanics of long-range interacting systems,
mean-field (MF) models are employed in many works (see \cite{Campa2009} and references therein).
In MF models, all the constituents interact equally with each other regardless of the distance.
It is expected that the MF models capture some qualitative features of general long-range interacting systems.
More strongly, some evidence of the {\it exactness of the MF theory} has been reported
for several models with the power-law interaction $1/r^{\alpha}$ with $0\leq\alpha< d$
~\cite{Cannas2000,Tamarit2000,Barre2002,Barre2005,Campa2000,Campa2003}.
Here, by the phrase ``exactness of the MF theory'' we mean that
the equilibrium properties of the system are equivalent to those of the corresponding MF model.
In other words, the free energy of the system is identical to the MF free energy.
Cannas {\it et al.} conjectured the exactness of the MF theory for the classical spin systems with long-range interactions~\cite{Cannas2000}
and several studies on specific models have followed it.

A previous work~\cite{Mori2010} demonstrated that
the exactness of the MF theory can be violated in a parameter region called the ``non-MF region'' for conserved systems.
Homogeneous configurations become unstable and a kind of phase transition should occur at the boundary between the MF region and the non-MF region.
In the van der Waals limit (see below), it is well known that the instability of the homogeneous states leads to
phase separation and the configuration becomes inhomogeneous.
These inhomogeneous states in the van der Waals limit can be considered as the coexistence of the two independent homogeneous phases, 
which are described by the MF theory.
On the other hand, when the interaction is nonadditive,
inhomogeneous states in the non-MF region cannot be described by the phase separation of two independent homogeneous phases,
and they are not described by the MF theory.
The aim of the present paper is to investigate the nature of this phase transition between the MF phase and the non-MF phase.

Although the previous work is concerned with pure ferromagnetic systems,
we also give an extension of the result to the spin glass systems.

The present work is organized as follows.
Details of the model and setting are given in Sec.~\ref{sec:setting}.
In Sec.~\ref{sec:result}, ``the exactness of the MF theory'' is examined
for nonconserved systems (systems where the magnetization is not conserved)
and for conserved systems (systems where the magnetization is conserved).
The exactness of the MF theory is always true for nonconserved systems
but not necessarily correct for conserved systems.
To demonstrate the above result, the Ising model with a long-range interaction is considered as an example in Sec.~\ref{sec:Ising}.
In Sec.~\ref{sec:phase}, the nature of the phase transition between the MF phase and the non-MF phase in conserved systems is investigated.
An application of our result to spin glasses is discussed in Sec.~\ref{sec:spin_glass}.
The summary of this paper is presented and some future problems are discussed in Sec.~\ref{sec:summary}.

\section{Setting}
\label{sec:setting}

We consider the following Hamiltonian on a $d$-dimensional lattice,
\beq
{\cal H}=-\frac{J}{2}\sum_{ij}^NK(\bm{r}_i-\bm{r}_j)\sigma_i\sigma_j-H\sum_i^N\sigma_i .
\label{eq:LR_H}
\eeq
Here we assumed the two-body long-range interaction.
We impose periodic boundary conditions and interpret the distance between the lattice points $i$ and $j$, $|\bm{r}_i-\bm{r}_j|$,
as the shortest distance of these lattice points in periodic boundary conditions.
The lattice interval is set to unity.
The parameter $J>0$ is a coupling strength and 
$K(\bm{r}_i-\bm{r}_j)$ is the interaction potential between the sites $i$ and $j$.
When $\sum_{\bm{r}_i\neq 0}K(\bm{r}_i)$ diverges,
the system is nonadditive and there exists no thermodynamic limit in the usual sense.
To avoid this difficulty, we normalize the interaction potential as
\beq
\sum_{\bm{r}_i\neq 0}K(\bm{r}_i)=1.
\label{eq:LR_K_nor}
\eeq
This is called Kac's prescription in the literature.
The ``spin'' variable $\sigma_i$ is arbitrary as long as it is finite;
in the Ising model $\sigma_i=\pm 1$,
in the classical $XY$ model $\sigma_i=(\cos\theta,\sin\theta)$,
and in the $q$-state Potts model $\sigma_i=\bm{e}_a$, $a=1,2,\dots,q$,
where $\bm{e}_a\cdot\bm{e}_b=\delta_{ab}$, and so on.
In this paper, we treat the one-component spin variable to make the presentation simple,
but the generalization to multicomponent spin variables is straightforward.

As the simplest long-range interacting model,
the infinite-range model exists,
\beq
{\cal H}_{\rm MF}=-\frac{J}{2N}\sum_{ij}^N\sigma_i\sigma_j-H\sum_i^N\sigma_i,
\eeq
for which it is known that the MF theory is exactly applicable.
Hereafter we call this model the ``MF model.''

In present paper, we consider the following two types of long-range interactions:
the power-law interaction
\beq
K(\bm{r})\propto\frac{1}{r^{\alpha}}, \quad 0\leq\alpha< d,
\label{eq:LR_power}
\eeq
and the Kac potential~\cite{Kac1963}
\beq
K(\bm{r})\propto\gamma^d\phi(\gamma\bm{r}).
\label{eq:LR_Kac}
\eeq
Here, $\phi(\bm{x})$ is assumed to be non-negative $\phi(\bm{x})\geq 0$ and integrable $\int d^dx\phi(\bm{x})<+\infty$.
Moreover, we assume that there is a positive and decreasing function $\psi(x)$ such that
\beq
\left\{
\begin{split}
|\nabla\phi(\bm{x})|\leq\psi(x), \\
\int d^dx\psi(x)<+\infty.
\end{split}
\right.
\label{eq:Kac_assumption}
\eeq
This assumption is necessary to justify the coarse-graining of the Hamiltonian discussed later.
A typical example of the Kac potential is the exponential form,
$K(\bm{r}_i-\bm{r}_j)\propto \gamma^de^{-\gamma |\bm{r}_i-\bm{r}_j|}$.
In this case, $\phi(\bm{x})=\psi(x)=e^{-x}$.

We will take the limit $\gamma\rightarrow 0$ in the Kac potential.
In this paper, two limiting procedures are considered:
the van der Waals limit $\gamma\rightarrow 0$ {\it after} $L\rightarrow\infty$~\cite{Lebowitz_Penrose1966}
and the long-range limit $\gamma\rightarrow 0$ with $\gamma L=\text{const.}$
The former limit corresponds to the situation where the interaction range $\gamma^{-1}$ is much longer than
the microscopic length scale (the lattice interval) but much shorter than the system size $L$.
In this case, the system is additive and it does not show anomalous behavior like the ensemble inequivalence.
The latter limit corresponds to the situation where the interaction range is comparable to the system size.
In this case, the system has no additivity.
These two limits give different behavior in general.


\section{Exactness of the MF theory}
\label{sec:result}

A condition of the exactness of the MF theory has been reported briefly~\cite{Mori2010}.
In this section, we give the detailed explanation of this property.

In the long-range interacting systems, it is expected that 
only long wavelength modes play important roles for macroscopic behavior.
In fact, it is possible to perform coarse graining {\it exactly} for long-range interacting models.

Now we explain what coarse graining is.
Let us divide the lattice system into blocks of the linear dimension $l$. 
The number of blocks is $(L/l)^d$ and each block has $l^d$ sites.
We introduce the local coarse-grained variable $m_p$ as 
\beq
m_p=\frac{1}{l^d}\sum_{i\in B_p}\sigma_i
\label{eq:LR_loc}
\eeq
in each block $B_p$, where $p=1,2,\dots ,(L/l)^d$.
We take the limit $L\rightarrow\infty$, $l\rightarrow\infty$ with $l/L\rightarrow 0$ (continuous limit).
This strategy is the same as the procedure in the paper by Barr\'e {\it et al.}~\cite{Barre2005}.
We define the position $\bm{x}_p=\bm{r}_p/L$, where $\bm{r}_p$ is the central position of a block 
$B_p$ [$p=1,\dots, (L/l)^d$].
We also define $m(\bm{x}_p)\equiv m_p$ .
For long-range interacting models, as shown in Appendix~\ref{sec:appendix}, the Hamiltonian is expressed only by $m(\bm{x})$ in the thermodynamic limit:
\beq
{\cal H}=\bar{{\cal H}}[m(\bm{x})]+o(N),
\label{eq:LR_cgH0}
\eeq
where
\begin{align}
\bar{{\cal H}}[m(\bm{x})]=&-\frac{NJ}{2}\int_{C_d}d^dx\int_{C_d}d^dy U(\bm{x}-\bm{y})m(\bm{x})m(\bm{y})
\nonumber \\
&-NH\int_{C_d}d^dx m(\bm{x}).
\label{eq:LR_cgH}
\end{align}
Here, the scaled interaction potential $U(\bm{x})$ is given by
\beq
U(\bm{x})=\lim_{L\rightarrow\infty}L^dK(L\bm{x}).
\label{eq:LR_cg_potential}
\eeq
The integrations in Eq.~(\ref{eq:LR_cgH}) are performed over a $d$-dimensional unit cube $C_d$, namely $\bm{x},\bm{y}\in [0,1]^d$.
Kac's prescription (\ref{eq:LR_K_nor}) implies
\beq
\int_{C_d}U(\bm{x})d^dx=1.
\label{eq:LR_U_nor}
\eeq

In the power-law potential and in the Kac potential with the long-range limit, 
$U(\bm{x})=\kappa_1/x^{\alpha}$ and $U(\bm{x})=\kappa_2\gamma_0^d\phi(\gamma_0\bm{x})$, respectively.
Here, $\kappa_1$ and $\kappa_2$ are normalization constants determined by Eq.~(\ref{eq:LR_U_nor}) and 
$\gamma_0=\gamma L$ is a constant in the long-range limit.
In the Kac potential with the van der Waals limit, $U(\bm{x})=\delta(\bm{x})$.

Performing the Fourier expansion in Eq.~(\ref{eq:LR_cgH}), we obtain the following expression:
\beq
\bar{{\cal H}}=-\frac{NJ}{2}\sum_{\bm{n}}U_{\bm{n}}|\hat{m}_{\bm{n}}|^2-NH\hat{m}_0,
\eeq
where
\begin{align}
m(\bm{x})&=\sum_{\bm{n}}\hat{m}_{\bm{n}}e^{2\pi i\bm{n}\cdot\bm{x}}, \\
U_{\bm{n}}&=\int_{C_d}d^dx U(\bm{x})\cos(2\pi\bm{n}\cdot\bm{x}).
\end{align}
We call $\{ U_{\bm{n}}\}$ {\it interaction eigenvalues}.
Interaction eigenvalues of ${\bm n}\neq 0$ are less than or equal to unity $U_{\bm n}\leq 1$,
because
\begin{align}
|U_{\bm n}|&=\left|\int_{C_d}d^dxU(\bm{x})\cos(2\pi\bm{n}\cdot\bm{x})\right| \nonumber \\
&\leq \int_{C_d}d^dx|U(\bm{x})|\cdot|\cos(2\pi\bm{n}\cdot\bm{x})| \nonumber \\
&\leq\int_{C_d}d^dxU(\bm{x})=1.
\end{align}

From now on, we consider the generalized free energy $F(m,T,H)$ which is defined as
\beq
\exp[-\beta F(m,T,H)]=
\sum_{[\{\sigma_i\}| \hat{m}_0=m]}e^{-\beta {\cal H}},
\label{eq:LR_free}
\eeq
where the summation is taken over the configurations with a fixed value of the total magnetization $\hat{m}_0=m$.
The temperature is $T=1/\beta$.
We can separate the long-wavelength modes from the short ones by coarse graining:
\beq
\sum_{[\{\sigma_i\}| \hat{m}_0=m]}
=\int_{\hat{m}_0=m}{\cal D}m(\bm{x})\sum_{\{\sigma_i\} \text{ with fixed $m(\bm{x})$}}.
\label{eq:LR_sum}
\eeq
The summation with the fixed $m(\bm{x})$ is expressed as
\beq
\sum_{\{\sigma_i\} \text{ with fixed $m(\bm{x})$}}1
=\exp\left(\int_{C_d}S(m(\bm{x}))d^dx\right),
\label{eq:LR_weight}
\eeq
where $S(m)$ is the {\it entropy}, 
\begin{align}
S(m)=\ln(&\text{the number of states} \nonumber\\ &\text{with the fixed magnetization $m$}).
\end{align}
From Eqs.~(\ref{eq:LR_cgH0}), (\ref{eq:LR_sum}), and (\ref{eq:LR_weight}), we obtain
\begin{align}
&\exp[-\beta F(m,T,H)]
=\int_{\hat{m}_0=m}{\cal D}m(\bm{x}) \nonumber \\
&\times\exp\left[-\beta \left(\bar{{\cal H}}[m(\bm{x})]-T\int_{C_d}S(m(\bm{x}))d^dx\right)\right].
\end{align}
By using the saddle-point method, we have
\begin{align}
F(m,T,H)&=\min_{\{m(\bm{x})|\hat{m}_0=m\}}
\left[\bar{\cal H}[m(\bm{x})]-T\int_{C_d}S(m(\bm{x}))d^dx\right] \nonumber \\
&\equiv \min_{\{m(\bm{x})|\hat{m}_0=m\}}{\cal F}(\{ m(\bm{x})\},T,H)
\label{eq:functional}
\end{align}
The functional ${\cal F}$ is called the {\it free energy functional} hereafter.
See Appendix~\ref{sec:saddle-point} for the rigorous justification of using the saddle-point method.

From Eq.~(\ref{eq:functional}) we obtain the upper bound
\begin{align}
F(m,T,H)&\leq{\cal F}[\{ m(\bm{x})=m\},T,H] \nonumber \\
&={\cal H}_{\rm MF}-TS(m) \nonumber \\
&=F_{\rm MF}(m,T,H).
\end{align}
The lower bound 
\beq
F(m,T,H)\geq F_{\rm MF}(m,T,H)-U_{\rm max}\Delta F_{\rm MF}(m,T/U_{\rm max},H)
\label{eq:low_bound}
\eeq
is obtained by replacing all the $U_{\bm{n}}$ with $\bm{n}\neq 0$ by $U_{\rm max}$ which is defined as
\beq
U_{\rm max}\equiv\max_{\bm{n}\neq 0}U_{\bm{n}}.
\eeq
The function $\Delta F_{\rm MF}$ is defined as
\beq
\Delta F_{\rm MF}\equiv F_{\rm MF}-{\rm CE}\{ F_{\rm MF}\},
\eeq
where ${\rm CE}$ means the convex envelope.
The convex envelope of a function $g(x)$ is defined as
the maximum convex function not exceeding $g(x)$ (see Fig.~\ref{fig:LR_convex_envelope}).
The derivation of the lower bound (\ref{eq:low_bound}) is given in Appendix~\ref{sec:low_bound}.

\begin{figure}[t]
\begin{center}
\includegraphics[scale=0.4]{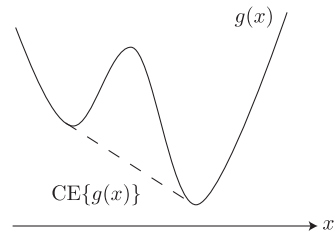}
\caption{An illustrative example of the convex envelope of the function $g(x)$.
The solid line denotes $g(x)$ and the dashed line denotes ${\rm CE}\{g(x)\}$.}
\label{fig:LR_convex_envelope}
\end{center}
\end{figure}

Collecting the upper and the lower bounds, we have
\begin{align}
&F_{\rm MF}(m,T,H)-U_{\rm max}\Delta F_{\rm MF}(m,T_{\rm eff},H)
\nonumber \\
&\leq F(m,T,H)\leq F_{\rm MF}(m,T,H),
\label{eq:LR_main}
\end{align}
where $T_{\rm eff}\equiv T/U_{\rm max}$.

An inequality necessary to prove Eq.~(\ref{eq:low_bound}) is Eq.~(\ref{eq:LR_Uapprox}) 
which corresponds to the replacement of $U_{\bm{n}}\rightarrow U_{\rm max}$ for all $\bm{n}\neq 0$.
In the Kac potential with the van der Waals limit, $U_{\bm n}=1$ for all $\bm{n}$.
Therefore, the equality $F(m,T,H)={\rm CE}\{F_{\rm MF}(m,T,H)\}$ holds for the Kac potential with the van der Waals limit.
It is a well known fact that the MF theory with the Maxwell's equal area law is justified in the van der Waals limit~\cite{Lebowitz_Penrose1966}.
The Maxwell's equal area law is equivalent to the replacement of the MF free energy by its convex envelope.
The replacement of the free energy by its convex envelope indicates the occurrence of the phase separation.

Next, we consider the local stability of the uniform configuration $m(\bm{x})=m$ for all $\bm{x}$.
If the uniform configuration gives the local maximum of the free energy functional,
there are configurations which have lower free energy than the MF.
Therefore, in this case, $F(m,T,H)<F_{\rm MF}(m,T,H)$ holds instead of $F(m,T,H)\leq F_{\rm MF}(m,T,H)$.
From this local stability analysis, we have
\begin{align}
F(m,T,H)&<F_{\rm MF}(m,T,H)
\nonumber \\
&\text{for $\frac{\d^2}{\d m^2}F_{\rm MF}(m,T_{\rm eff},H)<0$}.
\label{eq:LR_main2}
\end{align}

We have derived necessary inequalities for the generalized free energy with this.
According to Eqs.~(\ref{eq:LR_main}) and (\ref{eq:LR_main2}),
the parameter region $(m,T,H)$ is classified to the following three regions:
\begin{description}
\item[\it Region A]: the region where $\Delta F_{\rm MF}\left(m,T_{\rm eff},H\right)=0$.
In this region, the MF model gives the exact free energy, $F(m,T,H)=F_{\rm MF}(m,T,H)$.
\item[\it Region B]: the region where $\Delta F_{\rm MF}\left(m,T_{\rm eff},H\right)>0$ and 
$\frac{\d^2}{\d m^2}F_{\rm MF}\left(m,T_{\rm eff},H\right)\geq 0$.
In this region, it is not sure whether the MF model is exact, $F(m,T,H)\leq F_{\rm MF}(m,T,H)$.
However, the homogeneous states determined by the MF theory are locally stable.
\item[\it Region C]: the region where $\frac{\d^2}{\d m^2}F_{\rm MF}\left(m,T_{\rm eff},H\right)< 0$.
In this region, the MF model cannot describe the long-range interacting systems, $F(m,T,H)\neq F_{\rm MF}(m,T,H)$.
In this region, the homogeneous states are not even locally stable.
\end{description}

Notice that this classification is determined only by $F_{\rm MF}$ and $U_{\rm max}$.
Hence, we can specify these three regions concretely for individual models by analyzing only the MF models.

In region A and a part of B, $F(m,T,H)=F_{\rm MF}(m,T,H)$ holds.
This region is called the {\it MF region}.
On the other hand, in region C and the other part of B, $F(m,T,H)\neq F_{\rm MF}(m,T,H)$.
This region is called the {\it non-MF region}.

In region B, we cannot say whether a point $(m,T,H)$ belongs to a MF or non-MF region.
It depends on the type of the interaction, the value of the temperature, and the specific model (what $\sigma_i$ is).
However, the homogeneous state described by the MF theory is locally stable.

In the non-MF regin,
some inhomogeneous modes must develop.
Therefore, by observing the equilibrium configuration of the system with the conserved magnetization $m$,
we can know whether this point belongs to the MF or the non-MF region.
Namely, if the cluster appears in equilibrium, this point $(m,T,H)$ belongs to the non-MF region;
on the other hand, if the system is uniform, this point belongs to the MF region.

Here we discuss the exactness of the MF theory based on the derived inequalities.
In nonconserved systems, the equilibrium magnetization $m_{\rm eq}$ is determined by 
$\min_m[F(m,T,H)]=F(m_{\rm eq},T,H)$.
Because the equilibrium state belongs to the MF region where $\Delta F_{\rm MF}(m_{\rm eq},T_{\rm eff})=0$,
it is concluded that {\it exactness of the MF theory is true at any temperature in nonconserved systems}.

In contrast, in conserved systems, the generalized free energy itself is the equilibrium free energy and the derived inequalities mean that
the equilibrium property of the long-range interacting system is exactly the same 
as that of the corresponding MF model in the MF region.
On the other hand, they are not the same in the non-MF region.
As discussed above, the inhomogeneity appears in the non-MF region.
The clustering phenomena cannot be described by the MF model
with the help of the standard phase-separation argument.
Therefore, we conclude that
{\it exactness of the MF theory is violated in conserved systems}.
We investigate the clustering phenomena in Sec.~\ref{sec:phase}.

\section{Example: long-range interacting Ising model}
\label{sec:Ising}

\begin{figure}[tbhp]
\begin{center}
\includegraphics[clip,width=6cm]{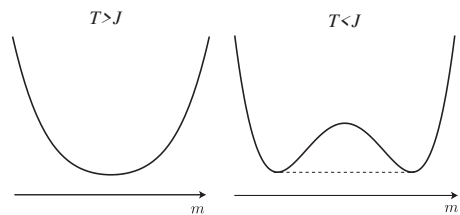}
\caption{The MF free energy of the Ising model (the solid line) and its convex envelope (the dashed line).
The left figure is for the case $T>J$.
As the MF free energy is convex for all $m$, it is equivalent to its convex envelope.
The right figure is for the case $T<J$.
In this case, the MF free energy is not convex, and the flat region appears in the convex envelope.}
\label{fig:LR_CE}
\end{center}
\end{figure} 

To make the statement clear, let us pick the long-range Ising model as a simple example.
We consider the Hamiltonian (\ref{eq:LR_H}) with $\sigma_i=\pm 1$.
The corresponding MF model is
\beq
{\cal H}_{\rm MF}=-\frac{J}{2N}\sum_{i\neq j}\sigma_i\sigma_j-H\sum_i\sigma_i\approx -\frac{NJ}{2}m^2-NHm.
\eeq
We can calculate the MF free energy, which is given by
\begin{align}
f_{\rm MF}=&-\frac{J}{2}m^2-Hm \nonumber \\
&+T\left(\frac{1+m}{2}\ln\frac{1+m}{2}+\frac{1-m}{2}\ln\frac{1-m}{2}\right).
\end{align}
In Fig.~\ref{fig:LR_CE},
the MF free energy and its convex envelope at $H=0$ are depicted.
For $T<T_c=J$, it becomes a nonconvex function of $m$.

When the ``effective temperature'' $T_{\rm eff}=T/U_{\rm max}$ is above the critical temperature, $T_c=J$, 
the relation $F_{\rm MF}(m,T_{\rm eff},H)={\rm CE}\{ F_{\rm MF}(m,T_{\rm eff},H)\}$ holds for any $m$.
In this case, Eq.~(\ref{eq:LR_main}) leads to
\beq
F(m,T,H)=F_{\rm MF}(m,T,H)
\eeq
for any $m$.
On the other hand, when $T_{\rm eff}<T_c$, there is a region where the MF free energy is not convex,
and from Eq.~(\ref{eq:LR_main}) we can conclude that
there is a region of $m$ such that $F(m,T,H)\neq F_{\rm MF}(m,T,H)$.

\begin{figure*}[t]
\begin{center}
\includegraphics[clip,width=10cm]{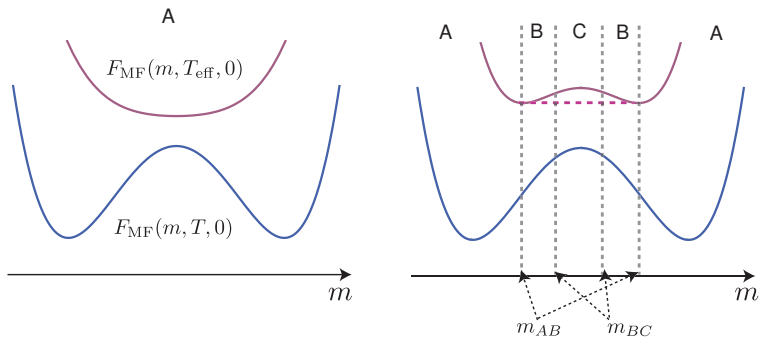}
\caption{(Color online)
An illustrative explanation of the relation of regions A, B, and C 
together with the MF free energy in the Ising model.
The upper line is the MF free energy at the temperature $T_{\rm eff}$
and the lower line is that at the genuine temperature $T$.
(a) The case of $T_{\rm eff}>T_c>T$.
(b) The case of $T_c>T_{\rm eff}>T$.}
\label{fig:LR_MF_region}
\end{center}
\end{figure*}

A schematic picture of the boundaries of three regions is depicted in Fig.~\ref{fig:LR_MF_region}.
The upper line is the MF free energy at the temperature $T_{\rm eff}=T/U_{\rm max}$
and the lower line is that at the genuine temperature $T$.
Figure~\ref{fig:LR_MF_region}(a) describes the case of $T_{\rm eff}>T_c>T$.
In this case, $F_{\rm MF}(m,T_{\rm eff},H)$ is a convex function of $m$.
Therefore, $\Delta F_{\rm MF}(m,T_{\rm eff},H)=0$ for any $m$ and $H$.
Figure~\ref{fig:LR_MF_region}(b) describes the case of $T_c>T_{\rm eff}>T$.
In this case, $F_{\rm MF}(m,T_{\rm eff},H)$ is not convex and 
deviation from the convex envelope (dashed line) appears.

In the Ising model with a long-range interaction, we can give the expression of regions A, B, and C explicitly.
In region A, $\Delta F_{\rm MF}(m,T_{\rm eff},H)=0$ holds.
Because $\Delta F_{\rm MF}$ is independent of $H$, the region A is given by 
\beq
|m|\geq |m_{\rm eq}(T_{\rm eff},H=0)|,
\eeq
where $m_{\rm eq}(T,H)$ is the magnetization in equilibrium which is given by the self-consistent equation,
$$m_{\rm eq}(T,H)=\tanh\left[\beta (Jm_{\rm eq}(T,H)+H)\right].$$

Similarly, region B is given by
\beq
|m_{\rm sp}(T_{\rm eff})|\leq |m|<|m_{\rm eq}(T_{\rm eff},0)|,
\eeq
where $m_{\rm sp}(T)$ is the spinodal point that is the metastability limit in the nonconserved systems,
\beq
m_{\rm sp}(T)=\pm\rt{1-\frac{1}{\beta J}}.
\eeq

Finally, region C is given by
\beq
|m|<|m_{\rm sp}(T_{\rm eff})|.
\eeq

After all, $m_{\rm AB}=m_{\rm eq}(T_{\rm eff},0)$ and $m_{\rm BC}=m_{\rm sp}(T_{\rm eff})$ in Fig.~\ref{fig:LR_MF_region}.

\begin{figure*}[t]
\begin{center}
\begin{tabular}{cc}
(a)&(b) \\
\includegraphics[clip,width=5cm]{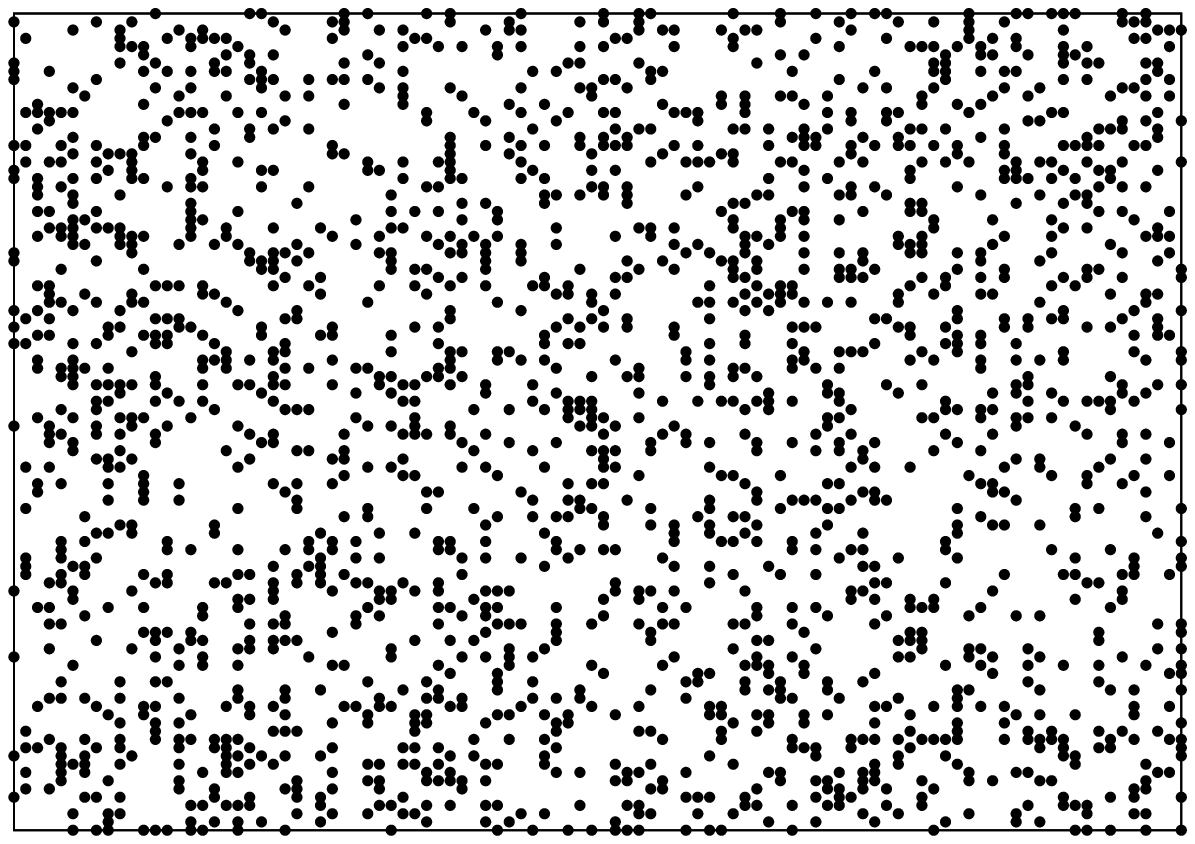}&
\includegraphics[clip,width=5cm]{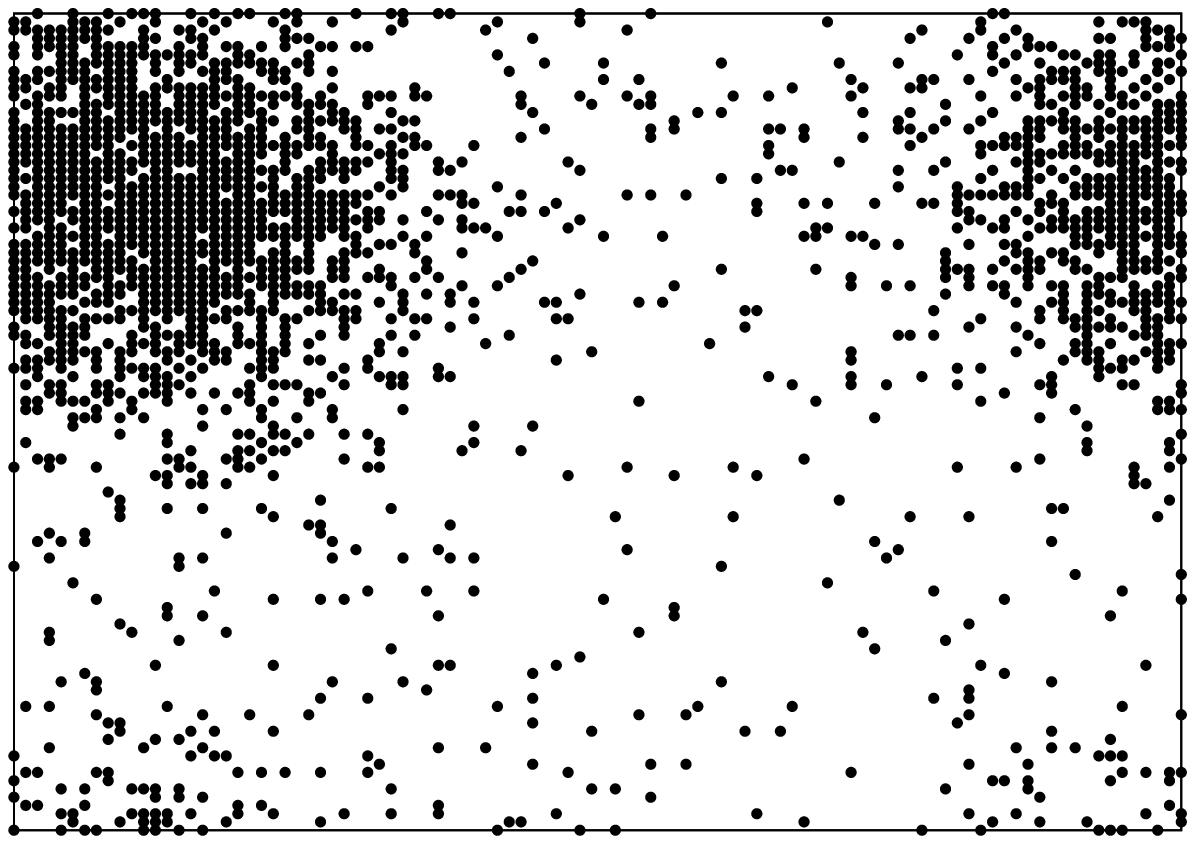}
\end{tabular}
\caption{Typical equilibrium snapshots in the two-dimensional (2D) lattice gas model (conserved Ising model)
 with $1/r$-type long-range interaction.
 Periodic boundary conditions are imposed.
 Black points represent occupied sites.
(a) Region A. The parameters are set to be $m=0.6$, $T=0.28$, $J=1$, and $L=100$.
(b) Region C. The parameters are set to be $m=0.6$, $T=0.18$, $J=1$, and $L=100$.}
\label{fig:LR_eq_cluster}
\end{center}
\end{figure*}

In conserved systems, a typical spin configuration is homogeneous at least in region A [see Fig.~\ref{fig:LR_eq_cluster}(a)]
but is inhomogeneous in region C [see Fig.~\ref{fig:LR_eq_cluster}(b)], as predicted in the previous section.

\begin{figure}[tbhp]
\begin{center}
\begin{tabular}{cc}
\raise20mm\hbox{$m$}&
\includegraphics[clip,width=7cm]{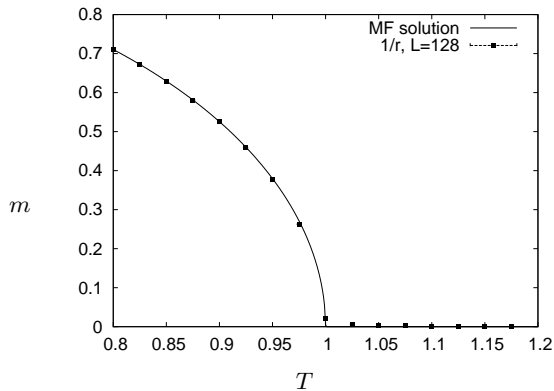}
\\ & $T$
\end{tabular}
\caption{Magnetizations against temperatures.
The points are numerical results of the 2D long-range Ising model with $K(r)\propto 1/r$ and $L=128$ calculated by the Monte Carlo method.
Error bars are smaller than the symbol size. 
The solid line is the MF solution.
They agree very well.}
\label{fig:mag}
\end{center}
\end{figure}

In nonconserved systems, we confirm the MF model is exact.
We depict spontaneous magnetizations for the two-dimensional Ising model with a long-range interaction ($\alpha=1$) in Fig.~\ref{fig:mag}.
As predicted in the previous section, it agrees with the MF result determined by solving the self-consistent equation, $m=\tanh\beta Jm$.
Moreover, it turns out that not only the equilibrium states 
but also the metastable states of the MF model given by the local minimum of the free energy are maintained,
because the local minimum is located in the range of $|m|\geq |m_{\rm sp}(T)|(\geq |m_{\rm sp}(T_{\rm eff})|)$; namely, it is located in region A or B.
In region A or B, homogeneous states are locally stable against the inhomogeneous fluctuations, 
and therefore the metastability defined in the MF model is not lost.
If they belonged to region C, they would have instability against inhomogeneous fluctuations and lose their local stability.

\section{Phase transition between MF phase and non-MF phase}
\label{sec:phase}

A kind of phase transition takes place between MF and non-MF regions in conserved systems.
We demonstrate a typical spin configuration of the Ising model with $1/r$ type long-range interaction in Fig.~\ref{fig:LR_eq_cluster}.
In Fig.~\ref{fig:LR_eq_cluster} (a), we depict a configuration at $T/J=0.28$ and $m=0.6$ which is in the MF region
and we find a homogeneous configuration.
In Fig.~\ref{fig:LR_eq_cluster} (b), we depict a configuration at $T/J=0.18$ and $m=0.6$ which is in the non-MF region,
where we find a clustered configuration.
We confirmed that there are two parameter regions, that is, the MF region and the non-MF region.

The next problem is to understand mechanism of the phase transition between the MF phase and the non-MF phase in conserved systems.
The Fourier modes with large interaction eigenvalues will play a significant role in the phase transition.

First consider the Kac potential with the van der Waals limit.
In this case, all the interaction eigenvalues are unity, $U_{\bm{n}}=1$;
all the Fourier modes contribute to the phase transition.
This fact indicates that the standard phase separation occurs and the system can be divided into two subsystems with different phases.

On the other hand, in the power-law interactions or the Kac potential with the long-range limit,
the spectrum of the interaction eigenvalues is discrete;
only the small number of Fourier modes which have the maximum interaction eigenvalue will contribute to the phase transition.
Therefore, it is expected that the phase transition is understood by the Landau expansion of the free energy functional by such Fourier modes.
It is also expected that this phase transition is qualitatively different from that in the Kac potential 
with the van der Waals limit,
which is well described by the notion of the phase separation.

We focus on the modes with the maximum interaction eigenvalue $U_{\rm max}$,
and the set of these modes are denoted by ${\cal M}$,
$${\cal M}\equiv \{ \bm{n}: \bm{n}\in \mathbb{Z}^d, U_{\bm{n}}=U_{\rm max}\}.$$
We regard $\hat{m}_{\bm{n}}$ with $\bm{n}\in {\cal M}$ as order parameters, and the other modes are determined by the condition 
\beq
\frac{\d{\cal F}(\{ m(\bm{x})\},T,H)}{\d m_{\bm{n}}}=0 \quad \bm{n}\notin {\cal M},
\label{eq:phase_cond}
\eeq
which is the equilibrium condition under the fixed order parameters.

In many cases, including the power-law interactions,
$U_{\bm{n}}$ tends to decrease as the length of the wave number vector $|\bm{n}|$ increases.
Therefore, we assume that
\beq
{\cal M}=\{\pm\bm{e}_{a}\}, \quad a=1,2,\dots, d
\label{eq:M}
\eeq
where $\bm{e}_a$ is a unit vector and $\bm{e}_a\cdot\bm{e}_b=\delta_{ab}$.
In this section, we assume that the interaction is isotropic.
Hence, the order parameters $\hat{m}_{\pm\bm{e}_a}$ do not depend on the direction $a$,
so we put $\phi\equiv \hat{m}_{\pm\bm{e}_a}$ and we regard $\phi$ as an order parameter.
Let us expand the free energy functional by $\{ \hat{m}_{\pm\bm{e}_a}\}$ with the help of Eq.~(\ref{eq:phase_cond}).
It is noticed that
\beq
\hat{m}_{\bm{n}}=O(|\phi|^{|n_1|+|n_2|+\dots+|n_d|}),
\eeq
which is derived from Eqs.~(\ref{eq:phase_cond}) and (\ref{eq:M}).
Up to the fourth order of $|\phi|$, we obtain the following expansion by some calculations:
\begin{widetext}
\begin{align}
{\cal F}=F_{\rm MF}+N\left\{
-d(JU_1+Ts^{(2)})|\phi|^2
+\left[\frac{d}{4}\frac{(Ts^{(3)})^2}{JU_2+Ts^{(2)}}+d(d-1)\frac{(Ts^{(3)})^2}{JU_{11}+Ts^{(2)}}
-\left(\frac{d}{4}+\frac{d(d-1)}{2}\right)Ts^{(4)}\right]|\phi|^4\right\}
\nonumber \\
+O(|\phi|^6).
\label{eq:Landau}
\end{align}
\end{widetext}
We defined $s^{(k)}\equiv d^ks(m)/dm^k$ and
\beq
\left\{
\begin{split}
U_1&\equiv U_{\pm\bm{e}_a}=U_{\rm max}, \\
U_{11}&\equiv U_{\bm{e}_a\pm \bm{e}_b} \quad a\neq b, \\
U_2&\equiv U_{\pm2\bm{e}_a}.
\end{split}
\right.
\eeq

Here we rewrite Eq.~(\ref{eq:Landau}) as
\beq
\frac{{\cal F}}{N}=f_{\rm MF}+a|\phi|^2+b|\phi|^4+O(|\phi|^6).
\eeq
It is noted that $a=0$ corresponds to $\d^2F_{\rm MF}(m,T_{\rm eff},H)/\d m^2=0$, that is,
the boundary between regions B and C.
It is evident that the first order phase transition occurs
if $b<0$ when $a=0$.
It is reasonably expected that the second-order transition occurs at the boundary of regions B and C
if $b>0$ when $a=0$,
though we do not show it strictly.
If we put $a=0$, then
\begin{align}
b=&\frac{dJU_1}{4}\left\{ (2d-1)\frac{s^{(4)}}{s^{(2)}}\right. \nonumber \\
&\left. -\left[4(d-1)\frac{U_1}{U_1-U_{11}}+\frac{U_1}{U_1-U_2}\right]\left(\frac{s^{(3)}}{s^{(2)}}\right)^2\right\}\nonumber \\
\equiv &b_0.
\label{eq:b0}
\end{align}
From the above discussion,
the transition is of the first order when $b_0<0$
and of the second order when $b_0>0$.

Let us demonstrate the above result in the Ising model ($\sigma_i=\pm 1$).
In this case, the entropy is given by
\beq
s(m)=-\frac{1+m}{2}\ln\frac{1+m}{2}-\frac{1-m}{2}\ln\frac{1-m}{2}.
\eeq
In this model,
\begin{align}
b_0\propto &(2d-1)(1+3m^2)\nonumber \\
&-2m^2\left[\frac{U_1}{U_1-U_2}+4(d-1)\frac{U_1}{U_1-U_{11}}\right] \nonumber \\
\equiv &(2d-1)(1+3m^2)-2m^2K.
\end{align}
If we put $b_0=0$,
\beq
m^2=\frac{2d-1}{2K-3(2d-1)}\equiv m_c^2.
\eeq
The first-order transition occurs when $|m|>|m_c|$ and the second-order transition occurs when $|m|<|m_c|$.
When $U_{11},U_2\rightarrow U_1$, then $b_0$ is negative for any $m\neq 0$.
Therefore, the transition is first order except for $m=0$ in the case where the spectrum of interaction eigenvalues is almost continuous.
On the other hand, if we put $U_{11}=U_2=0$, $b_0$ is positive and the transition is second order for any $m$.
Thus, we have found that the long-range nature of the interaction (the discreteness of the spectrum of interaction eigenvalues)
enhances the second order phase transition between homogeneous and inhomogeneous phases.

\begin{figure}[tbhp]
\begin{center}
\begin{tabular}{cc}
\raise20mm\hbox{$|\phi|$}&\includegraphics[clip,width=7.5cm]{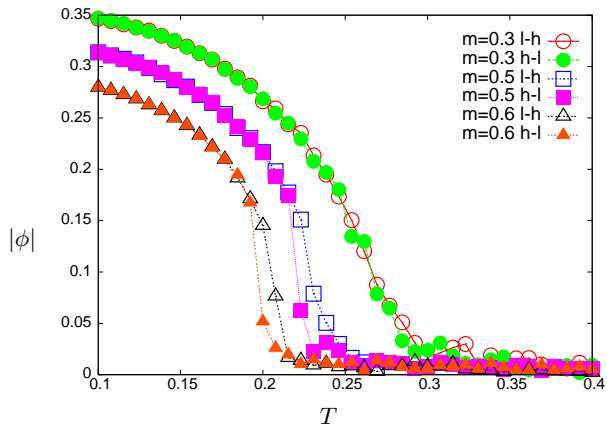}\\
&$T$
\end{tabular}
\caption{Monte Carlo results of average values of the order parameter $|\phi|$ 
with the temperature sweep from $T=0.1J$ to $0.4J$ (l-h) and from $0.4J$ to $0.1J$ (h-l).
The open circle is the data for $m=0.3$ and l-h, and the closed circle is for $m=0.3$ and h-l.
The open square is for $m=0.5$ and l-h, and the closed square is for $m=0.5$ and h-l.
The open triangle is for $m=0.6$ and l-h, and the closed triangle is for $m=0.6$ and h-l.
The system size is $L=60$ for $m=0.3$ and $L=100$ otherwise.}
\label{fig:transition}
\end{center}
\end{figure}

As an example, let us consider the Ising model on the two-dimensional square lattice with the interaction $K(r)\sim 1/r$.
In this case, $U_1\approx 0.310$, $U_{11}\approx 0.207$ and $U_2\approx 0.132$.
These parameters imply $m_c\approx 0.402$.
We demonstrate the result in Fig.~\ref{fig:transition}.
In this figure, we plot the average values of the order parameter $|\phi|$ in the Monte Carlo dynamics
under a temperature sweep over the range $T=0.1J$ to $0.4J$ and $T=0.4J$ to $0.1J$.
The transition is continuous when $m=0.3$ but it is discontinuous when $m=0.5$ and $m=0.6$.
Indeed, hysteresis loops appear for $m=0.5$ and $0.6$ as in Fig.~\ref{fig:transition}.

Thus, the phase transition between the MF and the non-MF phases does not necessarily occur just at $a=0$, the boundary of the regions B and C.
There are situations that the homogeneous states are locally stable but not globally stable.
This shows the relevance of considering the region B.

\section{Application to spin glasses}
\label{sec:spin_glass}

In this section, we discuss an application of the result discussed in Sec.~\ref{sec:result} to spin glass systems,
whose Hamiltonian is
\beq
{\cal H}=-\sum_{i<j}K_{ij}J_{ij}\sigma_i\sigma_j,
\label{eq:ham_glass}
\eeq
where $K_{ij}$ denotes the two-body interaction potential
and $J_{ij}$ is assumed to be a Gaussian random variable whose probability distribution $P(J_{ij})$ is given by
\beq
P(J_{ij})=\frac{1}{\rt{2\pi J^2}}\exp\left(-\frac{1}{2J^2}J_{ij}^2\right).
\eeq
The long-range spin glass systems have been extensively studied recently
in order to extend our knowledge of spin glasses in finite dimensions~\cite{Kotliar1983,Katzgraber2005,Leuzzi2009,Sharma2011}.
In these works, the universality class of the spin glass systems with $1/r^{\alpha}$-type interaction has been studied.
In one dimension, it has been revealed that the transition is in the MF universality class in the case of $1/2<\alpha<2/3$
and the non-MF universality class in the case of $2/3<\alpha<1$.
In this paper, we focus on the nonadditive regime, $\alpha\leq d/2$, and show that 
the system is fully identical to the MF model (Sherrington-Kirkpatrick model~\cite{Sherrington1975}) {\it at any temperature},
not only the critical exponents.

The free energy of this system is expressed as
\beq
f(\beta)=-\lim_{N\rightarrow\infty}\frac{1}{N\beta}\<\ln Z\>.
\eeq
Here the angular bracket denotes the average over the random interactions $\{ J_{ij}\}$
and $Z$ is the partition function,
$Z=\Tr e^{-\beta{\cal H}}$.
In this section, the symbol $\Tr$ refers to the summation over all the microscopic configurations $\{ \sigma_i\}$.

In order to examine the free energy, we apply the replica method, which is a nonrigorous but successful technique~\cite{Binder_rev1986}.
We can express the free energy as follows:
\beq
f(\beta)=-\lim_{N\rightarrow\infty}\frac{1}{N\beta}\lim_{n\rightarrow 0}\frac{\< Z^n\>-1}{n}.
\eeq
In the replica method, first we calculate $\< Z^n\>$ for integer $n$, then extrapolate it to noninteger $n$ and take the limit $n\rightarrow 0$.

Let us calculate $\< Z^n\>$ for integer $n$.
The quantity $\< Z^n\>$ is expressed as
\begin{widetext}
\beq
\< Z^n\>=\Tr\int\left(\prod \frac{dJ_{ij}}{\rt{2\pi J^2}}\right)\exp\left[-\frac{1}{2J^2}\sum_{i<j}J_{ij}^2
+\beta\sum_{a=1}^n\sum_{i<j}J_{ij}K_{ij}\sigma_i^a\sigma_j^a\right].
\eeq
\end{widetext}
By integrating out over $\{ J_{ij}\}$, we obtain
\beq
\< Z^n\>=\Tr\exp\left(\frac{1}{2}\beta^2J^2\sum_{a,b}^n\sum_{i<j}K_{ij}^2\sigma_i^a\sigma_i^b\sigma_j^a\sigma_j^b\right).
\eeq
If we define the vector $\vec{S}$ by
\beq
(\vec{S}_i)_{n(a-1)+b}\equiv\sigma_i^a\sigma_i^b,
\eeq
then we obtain the formal expression,
\beq
\< Z^n\>=\Tr\exp\left(\frac{1}{2}\beta^2J^2\sum_{i<j}K_{ij}^2\vec{S}_i\cdot\vec{S}_j\right).
\eeq
If we define the following quantities,
\beq
\tilde{\beta}\equiv\beta^2, \quad
\tilde{J}\equiv \frac{1}{2}J^2,\quad
\tilde{K}_{ij}\equiv K_{ij}^2,
\eeq
we obtain
\beq
\< Z^n\>=\Tr\exp\left(\tilde{\beta}\frac{\tilde{J}}{2}\sum_{ij}\tilde{K}_{ij}\vec{S}_i\cdot\vec{S}_j\right).
\eeq
In this form, we can apply the argument studied in the present paper on the exactness of the MF theory for pure ferromagnetic systems to spin glasses.
Namely, it is straightforward to show that MF theory is exact as long as the interaction $\tilde{K}_{ij}=K_{ij}^2$ is long range.
If we assume the power-law interaction $K_{ij}=C/r_{ij}^{\alpha}$,
then the MF model is exact when $\alpha<d/2$
instead of $\alpha<d$.
The constant $C$ is determined by the normalization condition
\beq
\sum_i\frac{C^2}{r_{ij}^{2\alpha}}=1,
\eeq
which is the correspondence of Eq.~(\ref{eq:LR_K_nor}).
(Note that in the infinite-range case, $C=1/\rt{N}$.)

From the above argument, we can conclude that $\< Z^n\>$ in the system (\ref{eq:ham_glass})
is exactly equal to that in the MF model at least for any integer $n$.
Therefore, the two free energies calculated by the replica method may be exactly equal.
Unfortunately, it is not certain if this result can be proved without using the replica method.
However, from the above result, it is reasonably expected that the true free energy of the system (\ref{eq:ham_glass})
is also exactly equal to that of the corresponding infinite-range model.
If the equilibrium properties of the system~(\ref{eq:ham_glass}) with $K_{ij}=C/r_{\ij}^{\alpha}$ and $\alpha<d/2$
in contact with a thermal reservoir at a temperature $T$ 
were different from those of the corresponding MF spin glasses,
then it would imply that the replica method is not exact at least for the long-range interacting spin glasses.
However, we have no reason to expect that the replica method does not work in long-range spin glasses,
and we conclude that spin glass systems with the Hamiltonian~(\ref{eq:ham_glass}) and $\alpha<d/2$
exhibits behavior identical with the corresponding MF models.

\section{Summary and Discussion}
\label{sec:summary}

We investigated the exactness of the MF theory in systems with nonadditive long-range interactions (power-law potential or the Kac potential with the long-range limit)
and additive long-range interactions (the Kac potential with the van der Waals limit) in a unified way.
We showed that the exactness of the MF theory is always valid for the nonconserved systems,
while for the conserved systems there exists the parameter region where the exactness of the MF theory is violated.
As an application of our result, we considered the spin glass system and revealed that
the exactness of the MF theory is valid also for long-range interacting spin glass systems within the treatment by the replica method,
as long as the square of the interaction potential is nonadditive.

We examined the nature of the phase transition between the MF phase and the non-MF phase by the Landau expansion of the free energy functional.
We pointed out that except for the van der Waals limit, inhomogeneous states observed in the non-MF region are quite different from
those created by the phase separation.
This aspect is reflected to the fact that only a small number of Fourier modes are important in the inhomogeneity. 
It is indicated by the discrete spectrum of interaction eigenvalues $U_{\bm{n}}$.
On the other hand, in the van der Waals limit, all the interaction eigenvalues are degenerate, $U_{\bm{n}}=1$, and the standard phase separation occurs.
It will be interesting to study dynamical nature associated with the phase transition between the MF phase and the non-MF phase,
because it is known that long-range interacting systems exhibit peculiar features also in dynamics.

The difference between conserved and nonconserved systems is a consequence of the violation of ensemble equivalence.
Whether the exactness of the MF theory holds or not depends on the specific ensemble.
A natural question is whether the MF theory is exact for a microcanonical ensemble.
It was investigated in Ref.~\cite{Barre2005}, where it turned out 
that the exactness of the MF theory is valid for the long-range Ising model in the microcanonical ensemble.
Here, it should be noted that in the MF Ising model, the canonical ensemble and the microcanonical ensemble are equivalent.
We can show from the result presented in Sec.~\ref{sec:result} 
that the exactness of the MF theory for long-range interacting systems is valid also in the microcanonical ensemble
{\it if the microcanonical and canonical ensembles are equivalent in the corresponding MF model}.
When the two ensembles are not equivalent in the MF model, however,
it is not obvious whether the exactness of the MF theory holds in the microcanonical ensemble.
This issue will be investigated elsewhere.

In non-conserved systems, the MF theory is always exact in equilibrium.
However, in the out-of-equilibrium situations, the inhomogeneity due to the long-range interactions may appear.
For example, let us consider the relaxation from the metastable states with the uniform magnetization profile.
The MF metastable states as local minimum of the free energy are also remained in the general long-range interacting systems,
as pointed out in Sec.~\ref{sec:Ising}.
However, at low temperatures these metastable states belong to region B.
If the metastable states belong to the non-MF region, these metastable states may relax to equilibrium by appearing the temporal inhomogeneity.
Hence, in these situations, the relaxation to equilibrium will be different from that observed in the MF models.

Recently, it was reported that a model of spin-crossover materials 
has an effective long-range interaction among molecules~\cite{Miyashita2009}.
In this model, although the Hamiltonian has only short-range interactions,
effective long-range interactions among molecules appear 
due to the lattice distortion by the difference of the molecular size depending on molecular states.
In such systems, the results of the MF model including the MF spinodal are not artifacts but physically relevant results.

In {\it small} systems, the range of the interaction can be of the order of the system size.
Indeed, negative heat capacity has been observed experimentally in small systems~\cite{Schmidt2001}.
For example, the phenomenon of the super-radiance originates from the effective long-range interaction among two-level atoms
mediated by the coupling with a cavity mode~\cite{Baumann2010}.
The similarity between long-range interacting macroscopic systems and small systems should be discussed in the future.

In this way, the nature of long-range interacting systems may be widely observed, and
it will become more important to study it from a general point of view.
 

\section*{Acknowledgements}
The author thanks Prof. S. Miyashita for numerous discussions, useful comments, and careful reading of the manuscript.
He also thanks Prof. A. P. Young for valuable comments on this work.
The Appendix.~\ref{sec:saddle-point} is owed to the fruitful discussion with Prof. Hal Tasaki.
The author acknowledges JSPS for financial support (Grant No. 227835).

\appendix
\section{Justification of the coarse-graining}
\label{sec:appendix}

We justify the coarse graining (\ref{eq:LR_cgH}) in this appendix.
The coarse-grained Hamiltonian with a finite system size $N=L^d$ and a finite number of blocks 
$\Omega=(L/l)^d$ is given by
\beq
\bar{{\cal H}}(N,\Omega)=-\frac{J}{2}\sum_{p,q}^{\Omega}U_{pq}m_pm_q-Hl^d\sum_p^{\Omega}m_p,
\label{eq:Hbar}
\eeq
where
\beq
U_{pq}=\sum_{i\in B_p}\sum_{j\in B_q}K(\bm{r}_{ij}), \quad \bm{r}_{ij}=\bm{r}_i-\bm{r}_j,
\eeq
and $m_p$ is the average global variable of the block $B_p$ defined by Eq.~(\ref{eq:LR_loc}).
Notice that $\bar{{\cal H}}(N,\Omega)$ approaches the coarse-grained Hamiltonian (\ref{eq:LR_cgH})
when the limit $N\rightarrow\infty$, $\Omega\rightarrow\infty$ with $\Omega /N\rightarrow 0$ is taken.

Our aim is to prove that there exists a sequence $\Omega(N)$ such that
\beq
\lim_{N\rightarrow\infty}\frac{1}{N}\sup_{\{\sigma_i\}}\left|
{\cal H}-\bar{{\cal H}}(N,\Omega(N))\right|=0,
\eeq
and satisfies
\beq
\left\{
\begin{split}
\lim_{N\rightarrow\infty}\Omega(N)=\infty, \\
\lim_{N\rightarrow\infty}\frac{\Omega(N)}{N}=0.
\end{split}
\right.
\label{eq:K_condition}
\eeq
When we consider the Kac potential with the van der Waals limit,
we replace $\lim_{N\rightarrow\infty}$ and $\Omega(N)$ by $\lim_{\gamma\rightarrow 0}\lim_{N\rightarrow\infty}$
and $\Omega(N,\gamma)$, respectively.

The following derivation is a strightforward generalization of the strategy of the paper by Barr\'e {\it et al.}~\cite{Barre2005},
which they proved only for the one-dimensional Ising model with the power-law interaction.

\begin{figure}[t]
\begin{center}
\includegraphics[scale=0.4]{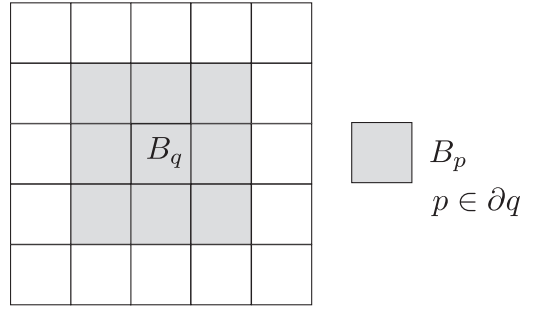}
\caption{An illustrative explanation of the set $\d q$ for $d=2$.
The central block is $B_q$ and the surrounding gray blocks are in $\d q$.}
\label{fig:LR_block}
\end{center}
\end{figure}

First, we express the coarse-grained Hamiltonian in terms of the microscopic variables $\{\sigma_i\}$.
Then
\beq
\bar{{\cal H}}(N,\Omega)=-\frac{J}{2}\sum_{p,q}^{\Omega}U_{pq}\frac{1}{l^{2d}}\sum_{i\in B_p, j\in B_q}\sigma_i\sigma_j-H\sum_i\sigma_i,
\eeq
where $l$ is the linear dimension of a block.
We thereby have
\begin{align}
&\frac{1}{N}|{\cal H}-\bar{{\cal H}}| \nonumber \\
&=\frac{J}{2Nl^{2d}}\left|
\sum_{p,q}^{\Omega}\sum_{i\in B_p, j\in B_q}\sigma_i\sigma_j\sum_{k\in B_p,l\in B_q}[K(\bm{r}_{ij})-K(\bm{r}_{kl})]\right|
\nonumber \\
&\leq \frac{J}{2Nl^{2d}}\sum_{p,q}^{\Omega}\sum_{i\in B_p, j\in B_q}|\sigma_i\sigma_j|\sum_{k\in B_p,l\in B_q}|K(\bm{r}_{ij})-K(\bm{r}_{kl})|.
\end{align}
Assuming $|\sigma_i\sigma_j|\leq C$, we have
\begin{align}
&\frac{1}{N}|{\cal H}-\bar{{\cal H}}|\nonumber \\
&\leq \frac{CJ}{2Nl^{2d}}\sum_{p,q}^{\Omega}\sum_{i\in B_p, j\in B_q}
\sum_{k\in B_p,l\in B_q} |K(\bm{r}_{ij})-K(\bm{r}_{kl})|.
\label{eq:LR_dev}
\end{align}
We define $D_{pq}$ as the shortest distance between the two blocks $B_p$ and $B_q$.
Namely, 
\beq
D_{pq}\equiv \min_{k\in B_p, l\in B_q}|\bm{r}_k-\bm{r}_l|.
\eeq
Moreover, the set $\d q$ is defined as the set of $p$ such that $D_{pq}<l$ (see Fig.~\ref{fig:LR_block}).
We divide the summation of Eq.~(\ref{eq:LR_dev}) into two terms:
the term with $p\in \d q$ and the term with $p\notin \d q$.
Then we obtain
\begin{align}
&\frac{1}{N}|{\cal H}-\bar{{\cal H}}|\nonumber \\
&\leq \frac{CJ}{2Nl^{2d}}\sum_{q=1}^{\Omega}\sum_{p\in \d q}\sum_{i\in B_p, j\in B_q}\sum_{k\in B_p,l\in B_q} |K(\bm{r}_{ij})-K(\bm{r}_{kl})|
\nonumber \\
&+\frac{CJ}{2Nl^{2d}}\sum_{q=1}^{\Omega}\sum_{p\notin \d q}\sum_{i\in B_p, j\in B_q}\sum_{k\in B_p,l\in B_q} |K(\bm{r}_{ij})-K(\bm{r}_{kl})|
\nonumber \\
&\equiv A_1+A_2.
\end{align}

From now on, we prove that there exists a sequence $\Omega(N)$ such that
$\lim_{N\rightarrow\infty}A_1=\lim_{N\rightarrow\infty}A_2=0$
both for the power-law potential and the Kac potential.

\subsection*{Upper bound of $A_1$}

If we define $K_0$ as the maximum value of $|K(r_{ij})|$, then
\beq
|K(\bm{r}_{ij})-K(\bm{r}_{kl})|\leq |K(\bm{r}_{ij})|+|K(\bm{r}_{kl})|\leq 2K_0.
\eeq
Therefore, we obtain
\beq
A_1\leq \frac{CJ}{2Nl^{2d}}\sum_{q=1}^{\Omega}\sum_{p\in \d q}
l^{4d}\cdot 2K_0 =\frac{CJK_0l^{2d}}{N}\sum_{q=1}^{\Omega}\sum_{p\in \d q}1,
\eeq
where we used $\sum_{i\in B_p}1=l^d$.
Because the number of blocks which belong to $\d q$ is determined by the spatial dimension $d$,
we can write
\beq
\sum_{p\in \d q}1\equiv n_d.
\eeq
Then we obtain the upper bound of $A_1$:
\beq
A_1\leq \frac{CJK_0l^{2d}}{N}\Omega n_d.
\eeq
As $\Omega=(L/l)^d$ and $N=L^d$,
\beq
A_1\leq Cn_dK_0l^d.
\eeq
Remember that $K_0$ depends on the system size $L$ or the interaction length $\gamma$,
because we normalize the interaction potential so that 
\beq
\sum_{\bm{r}_i\neq 0}K(\bm{r}_i)=1.
\eeq

In the case of the Kac potential, $K(\bm{r})\sim \gamma^d \phi(\gamma \bm{r})$.
Here, the symbol $\sim$ means equal except for a nonessential factor independent of $l$, $L$, and $\gamma$.
Therefore, the symbol $\sim$ does not imply any approximations (we use the symbol $\lesssim$ similarly).
Thus $K_0\sim \gamma^d$ and 
\beq
K_0l^d\sim (\gamma l)^d.
\eeq
In the van der Waals limit, we take $\gamma\rightarrow 0$ after $L\rightarrow\infty$.
Hence if we take the limit $l\rightarrow\infty$ {\it after} the limit $\gamma\rightarrow 0$,
then $A_1\rightarrow 0$ and the conditions (\ref{eq:K_condition}) are satisfied.

In the long-range limit, where $L\rightarrow\infty$ with $\gamma L={\rm const.}$ is taken,
if we take $l\rightarrow\infty$ after $L\rightarrow\infty$, then
$A_1\rightarrow 0$ and the conditions (\ref{eq:K_condition}) are also satisfied.

In the case of the power-law potential, $K(r)\sim L^{\alpha-d}/r^{\alpha}$ and $K_0\sim L^{\alpha -d}$ for $0\leq\alpha <d$.
Therefore,
\beq
A_1\sim K_0l^d\sim \left(\frac{l}{L^{1-\alpha/d}}\right)^d.
\eeq
After all, $A_1\rightarrow 0$ when we take the limit of $l\rightarrow\infty$ after $L\rightarrow\infty$.
When $\alpha> d$, the function $l(L)$ which satisfies Eq.~(\ref{eq:K_condition}) does not exist, and the coarse graining cannot be performed exactly.

\subsection*{Upper bound of $A_2$}

As $p\notin \d q$, $D_{pq}\neq 0$ and the following inequality is satisfied for $\bm{r}_i, \bm{r}_k\in B_p$
and $\bm{r}_j, \bm{r}_l\in B_q$:
\begin{align}
|\bm{r}_{ij}-\bm{r}_{kl}|&=\left| (\bm{r}_i-\bm{r}_k)-(\bm{r}_j-\bm{r}_l)\right| \nonumber \\
&\leq |\bm{r}_i-\bm{r}_k|+|\bm{r}_j-\bm{r}_l| \nonumber \\
&=r_{ik}+r_{jl} \nonumber \\
&\leq 2\rt{d}l\equiv Dl
\end{align}
From the mean-value theorem, there exists $0\leq \theta\leq 1$ such that
\beq
K(\bm{r}_{ij})-K(\bm{r}_{kl})=[\nabla K(\bm{r}_{ij}+\theta(\bm{r}_{kl}-\bm{r}_{ij}))]\cdot(\bm{r}_{kl}-\bm{r}_{ij}).
\eeq
Combining the above two relations, we have
\beq
|K(\bm{r}_{ij})-K(\bm{r}_{kl})|\leq Dl|\nabla K(\bm{r}_{ij}+\theta(\bm{r}_{kl}-\bm{r}_{ij}))|.
\eeq
Here, for $p\neq q$,
\begin{align}
&|\nabla K(\bm{r}_{ij}+\theta(\bm{r}_{kl}-\bm{r}_{ij}))|\nonumber \\
&\leq \max_{0\leq\theta\leq 1}
\max_{\bm{r}_i,\bm{r}_k\in B_p, \bm{r}_j,\bm{r}_l\in B_q}|\nabla K(\bm{r}_{ij}+\theta(\bm{r}_{kl}-\bm{r}_{ij}))|\nonumber \\
&\equiv K'_{pq}.
\end{align}
Using these inequalities, we can evaluate the upper bound of $A_2$:
\beq
A_2\leq \frac{CDJl^{2d+1}}{2N}\sum_{q=1}^{\Omega}\sum_{p\notin \d q} K'_{pq}.
\label{eq:A2inequality}
\eeq

\begin{figure}[t]
\begin{center}
\includegraphics[scale=0.4]{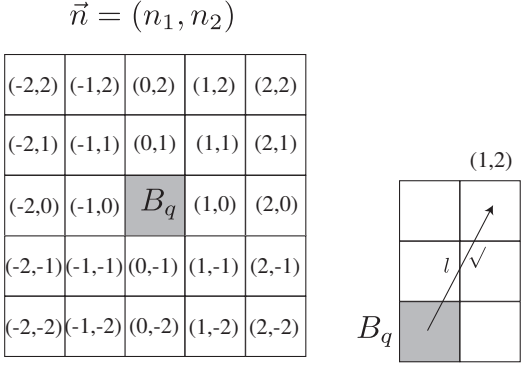}
\caption{An illustrative explanation of $\vec{n}$ for $d=2$.
The distance between the central positions of $B_{\vec{n}}$ and $B_q$ is given by $l|\vec{n}|$.
Moreover, for all $\vec{n}$, $B_{\vec{n}}\notin \d q$,
there exists a constant $0<a<1$ such that
the shortest distance between $B_{\vec{n}}$ and $B_q=B_0$ is restricted by $D_{\vec{n},0}\geq al|\vec{n}|$.
For $d=2$, this constant is $a=1/\sqrt{5}$ (see the right figure).}
\label{fig:LR_block2}
\end{center}
\end{figure}

As the blocks are aligned in the $d$-dimensional space, each block can be labeled by the $d$-dimensional vector
$$\vec{n}=(n_1,n_2,\cdots,n_d),$$
where $n_i$ is an integer and $-L/2l\leq n_i\leq L/2l$.
We fix the block $B_q$ at $\vec{n}=0$, and we determine $\vec{n}$ so that
the configuration $\bm{r}_p$ of the center of a block $B_p$ may be given by $\bm{r}_p=l\vec{n}$.
In this case, there is a constant $0<a<1$ such that $D_{pq}\geq al|\vec{n}|$ 
for all $p$ with $B_p\notin \d q$ (see Fig.~\ref{fig:LR_block2}).  
When we assume periodic boundary conditions,\footnote
{Periodic boundary conditions are not necessary to prove $A_2\rightarrow 0$.
It is only for convenience.}
the summation can be written as follows:
\begin{align}
\sum_{q=1}^{\Omega}\sum_{p\notin \d q}K_{pq}&\leq \Omega\sum_{n_1=-L/2l}^{L/2l}\sum_{n_2=-L/2l}^{L/2l}
\cdots\sum_{n_d=-L/2l}^{L/2l}K'_{\vec{n},0}
\nonumber \\
&\equiv \Omega\sum_{\vec{n}}K'_{\vec{n},0},
\end{align}
where the index $p$ corresponds to the vector $\vec{n}$ and we write $K'_{pq}$ as $K'_{\vec{n},0}$
as well as we defined $K'_{\vec{n},0}=0$ for $\vec{n}\in\d q$.

Thus, from (\ref{eq:A2inequality}), the upper bound of $A_2$ is given by
\beq
A_2\leq \frac{CDJ}{2}l^{d+1}\sum_{\vec{n}}K'_{\vec{n},0}.
\eeq
Next, we consider each case of the interaction forms and evaluate the upper bound of $A_2$.

In the Kac potential, $K(r)\sim \gamma^d\phi(\gamma r)$ and therefore 
\beq
K'(r)=\gamma^{d+1}\phi'(\gamma r).
\eeq
From the assumption of the Kac potential (\ref{eq:Kac_assumption}),
\beq
K'_{\vec{n},0}=\gamma^{d+1}\psi(D_{\vec{n},0})\leq\gamma^{d+1}\psi(al|\vec{n}|).
\eeq
Therefore, we obtain the upper bound of $A_2$,
\begin{align}
A_2\lesssim l^{d+1}\sum_{\vec{n}}K'_{\vec{n},0}
&\leq(\gamma l)^{d+1}\sum_{\vec{n}}\psi(a\gamma l|\vec{n}|) \nonumber \\
&\approx(\gamma l)^{d+1}\frac{1}{(a\gamma l)^d}\int_{-\infty}^{\infty}d^dx\psi(x) \nonumber \\
&\sim\gamma l
\end{align}
This tends to 0 when we take the limit $l\rightarrow\infty$ after $\gamma\rightarrow 0$.
Hence we can conclude that $A_2\rightarrow 0$.

Next, we consider the power-law potential.
The power-law potential is expressed as $K(r)\sim L^{\alpha-d}/r^{\alpha}$ for $0\leq\alpha<d$ and its derivative is
\beq
K'(r)\sim -\alpha \frac{L^{\alpha-d}}{r^{\alpha+1}}.
\eeq
We can evaluate $K'_{\vec{n},0}$ as
\beq
K'_{\vec{n},0}\sim \frac{L^{\alpha-d}}{D_{\vec{n},0}^{\alpha+1}}
\leq\frac{L^{\alpha-d}}{(al|\vec{n}|)^{\alpha+1}}.
\eeq
Therefore, when $0\leq\alpha<d-1$,
\beq
A_2\lesssim l^{d+1}\sum_{\vec{n}}K'_{\vec{n},0}
\lesssim l^{d+1}\frac{L^{\alpha-d}}{l^{\alpha+1}}\left(\frac{L}{l}\right)^{d-\alpha-1}
=\frac{l}{L}.
\eeq
When $\alpha=d-1$,
\beq
A_2\lesssim l^{d+1}\frac{L^{-1}}{l^d}\sum_{\vec{n}}\frac{1}{|\vec{n}|^d}
\sim \frac{l}{L}\ln\left(\frac{L}{l}\right).
\eeq
When $d-1<\alpha<d$, since $\sum_{\vec{n}}(1/|\vec{n}|^{\alpha+1})$ is finite in the limit $L/l\rightarrow\infty$,
\beq
A_2\lesssim l^{d+1}\sum_{\vec{n}}K'_{\vec{n},0}\lesssim l^{d+1}\frac{L^{\alpha-d}}{l^{\alpha+1}}
=\left(\frac{l}{L}\right)^{d-\alpha}.
\eeq
In any case, we have $A_2\rightarrow 0$ when the conditions (\ref{eq:K_condition}) are fulfilled.

In this way, we proved (\ref{eq:LR_CE}) both for the Kac potential and the power-law potential.

\section{The rigorous justification of the saddle-point method}
\label{sec:saddle-point}

In Appendix.~\ref{sec:appendix}, we proved that there exists a function $g(L,l)$ such that
\beq
|{\cal H}-\bar{{\cal H}}|\leq L^dg(L,l)
\label{eq:g1}
\eeq
and
\beq
\lim_{l\rightarrow\infty}\lim_{L\rightarrow\infty}g(L,l)=0.
\label{eq:g2}
\eeq
For the definition of $\bar{\cal H}$, see Eq.~(\ref{eq:Hbar}).
We evaluate the partition function
\beq
Z=\sum_{\{\sigma_i\}}{}^{'}e^{-\beta{\cal H}},
\eeq
where $\sum{}^{'}$ denotes the summation over the spin configuration $\{\sigma_i\}$ under the restriction of $\frac{1}{N}\sum_i\sigma_i=m$.
From Eq.~(\ref{eq:g1}), the partition function is bounded by
\begin{align}
\sum_{\{ m_p\}}{}^{'}\exp\left[-\beta\bar{{\cal H}}+\sum_p^{\Omega}S(m_p)-\beta L^dg(L,l)\right] \nonumber \\
\leq Z\leq \sum_{\{ m_p\}}{}^{'}\exp\left[-\beta\bar{\cal H}+\sum_p^{\Omega}S(m_p)+\beta L^dg(L,l)\right].
\end{align}
Here, $\sum_{\{ m_p\}}{}^{'}$ denotes the summation over all possible values of $\{ m_p\}$ 
with a restriction of $\frac{1}{\Omega}\sum_{p}^{\Omega}m_p=m$.
We defined $S(m_p)\equiv \ln \sum_{\{\{\sigma_i\}:i\in B_p, \frac{1}{l^d}\sum_{i\in B_p}\sigma_i=m_p\}}1$.
Let us define $$-\beta F^*\equiv \max_{\{ m_p\}:\frac{1}{\Omega}\sum_p^{\Omega}m_p=m}\left[-\beta\bar{\cal H}+\sum_p^{\Omega}S(m_p)\right].$$
Then, a lower bound of $Z$ is given by
\beq
Z\geq e^{-\beta F^*-\beta L^dg(L,l)}.
\label{eq:Z_lower}
\eeq
An upper bound is also obtained as follows:
\begin{align}
Z&\leq \sum_{\{ m_p\}}{}^{'}e^{-\beta F^*+\beta L^dg(L,l)} \nonumber \\
&\leq \sum_{\{ m_p\}}e^{-\beta F^*+\beta L^dg(L,l)}.
\end{align}
Here, we assume that $\sum_{m_p}1\leq cl^{nd}$ with a positive constant $c$ and a positive integer $n$, both of which are independent of $L$ and $l$.
We can confirm that this assumption is satisfied in individual models.
For example, in the Ising model, $m_p$ takes the following value
$$l^dm_p\in\{ l^d,l^d-2,l^d-4,\dots,-l^d\}.$$
Therefore, in this case, $\sum_{m_p}1=l^d+1\leq 2l^d$.
Also in other models, we can check this property.
Therefore, we have
\beq
Z\leq (cl^{nd})^{\Omega}e^{-\beta F^*+\beta L^dg(L,l)}.
\label{eq:Z_upper}
\eeq
(note that $\Omega=(L/l)^d$.)
From Eqs.~(\ref{eq:Z_lower}) and (\ref{eq:Z_upper}), we have
\beq
\frac{F^*}{L^d}-g(L,l)-\frac{1}{\beta}\frac{\ln(cl^{nd})}{l^d}
\leq -\frac{1}{\beta}\frac{1}{L^d}\ln Z\leq \frac{F^*}{L^d}+g(L,l).
\eeq
Hence, using Eq.~(\ref{eq:g2}), we obtain
\begin{align}
f&\equiv -\frac{1}{\beta}\lim_{l\rightarrow\infty}\lim_{L\rightarrow\infty}\frac{1}{L^d}\ln Z \nonumber \\
&=\lim_{l\rightarrow\infty}\lim_{L\rightarrow\infty}\frac{F^*}{L^d}.
\end{align}
Thus, the saddle-point method in Eq.~(\ref{eq:functional}) is justified rigorously.

\section{The derivation of the lower bound (\ref{eq:low_bound})}
\label{sec:low_bound}

From Eq.~(\ref{eq:functional}) and $F_{\rm MF}(m,T,H)=\tilde{\cal H}[m]-TS(m)$, we have
\begin{align}
&F(m,T,H)-F_{\rm MF}(m,T,H) \nonumber \\
&=\min_{\{m(\bm{x})|\hat{m}_0=m\}}\left[\tilde{\cal H}-T\int_{C_d}S(m(\bm{x}))d^dx\right]+TS(m)
\nonumber \\
&=\min_{\{m(\bm{x})|\hat{m}_0=m\}}\left[
-\frac{NJ}{2}\sum_{\bm{n}\neq 0}U_{\bm{n}}|\hat{m}_{\bm{n}}|^2
-T\int_{C_d}S(m(\bm{x}))d^dx\right] \nonumber \\
&\hspace{2cm}+TS(m).
\label{eq:LR_var}
\end{align}
Let us define $U_{\rm max}$ as the maximum interaction eigenvalue except for $U_0=1$,
\beq
U_{\rm max}\equiv \max_{\bm{n}\neq 0}U_{\bm{n}}.
\eeq
Taking into account of the relations
\beq
\sum_{\bm{n}\neq 0}U_{\bm{n}}|\hat{m}_{\bm{n}}|^2
\leq U_{\rm max}\sum_{\bm{n}\neq 0}|\hat{m}_{\bm{n}}|^2,
\label{eq:LR_Uapprox}
\eeq
and
\beq
\int_{C_d}m(\bm{x})^2d^dx=m^2+\sum_{\bm{n}\neq 0}|\hat{m}_{\bm{n}}|^2,
\eeq
the RHS of (\ref{eq:LR_var}) is written as
\begin{align}
&-\frac{NJ}{2}\sum_{\bm{n}\neq 0} U_{\bm{n}}|\hat{m}_{\bm{n}}|^2
-T\int_{C_d}\left(S(m(\bm{x}))-S(m)\right)d^dx
\nonumber \\
\geq &\frac{NJ}{2}U_{\rm max}m^2+TS(m) \nonumber \\
&+\int_{C_d}d^dx\left[-\frac{NJ}{2}U_{\rm max}m(\bm{x})^2-TS(m(\bm{x}))\right]
\nonumber \\
=&-U_{\rm max}\bigg[F_{\rm MF}(m,T/U_{\rm max},H) \nonumber \\
&\ \ \left.-\int_{C_d}d^dx F_{\rm MF}(m(\bm{x}),T/U_{\rm max},H)\right].
\label{eq:LR_inequality}
\end{align}
Thus, we obtain
\begin{align}
&F(m,T,H)\geq F_{\rm MF}(m,T,H) \nonumber \\
&+U_{\rm max}\min_{\{m(\bm{x})|\hat{m}_0=m\}}
\left[\int_{C_d}d^dx F_{\rm MF}(m(\bm{x}),T/U_{\rm max},H)\right]
\nonumber \\
&-U_{\rm max}F_{\rm MF}(m,T/U_{\rm max},H).
\end{align}

Here, we can prove
\begin{align}
\min_{\{m(\bm{x})|\hat{m}_0=m\}}\left[\int_{C_d}d^dx F_{\rm MF}(m(\bm{x}),T/U_{\rm max},H)\right]
\nonumber \\
={\rm CE}\{ F_{\rm MF}(m,T/U_{\rm max},H)\}.
\label{eq:LR_CE}
\end{align}
Therefore, we obtained the lower bound (\ref{eq:low_bound}).

Let us show Eq.~(\ref{eq:LR_CE}).
Later, we omit the temperature dependence and the external field dependence of the free energy.
Let us define a function $G(m)$ as
\beq
G(m)\equiv\min_{\{m(\bm{x})|\hat{m}_0=m\}}
\left[\int_{C_d}d^dx F_{\rm MF}(m(\bm{x}))\right].
\eeq
If we consider the uniform configuration, $m(\bm{x})=m$, we have
$\int_{C_d}d^dxF_{\rm MF}(m(\bm{x}))=F_{\rm MF}(m)$.
Therefore we obtain $G(m)\leq F_{\rm MF}(m)$.
From the definition of the convex envelope,
\begin{align}
G(m)&=\min_{\{m(\bm{x})|\hat{m}_0=m\}}
\left[\int_{C_d}d^dx F_{\rm MF}(m(\bm{x}))\right]
\nonumber \\
&\geq\min_{\{m(\bm{x})|\hat{m}_0=m\}}
\left[\int_{C_d}d^dx {\rm CE}\left\{F_{\rm MF}(m(\bm{x}))\right\}\right]\nonumber \\
&\geq\min_{\{m(\bm{x})|\hat{m}_0=m\}}
\left[{\rm CE}\left\{F_{\rm MF}\left(\int_{C_d}d^dx m(\bm{x})\right)\right\}\right] \nonumber \\
&={\rm CE}\{F_{\rm MF}(m)\}.
\end{align}
Hence, we obtain
\beq
{\rm CE}\{F_{\rm MF}(m)\}\leq G(m)\leq F_{\rm MF}(m).
\label{eq:G_inequality}
\eeq

Next let us prove that $G(m)$ is a convex function of $m$.
For a constant $0<\lambda<1$, we consider
\beq
\lambda G(m_1)+(1-\lambda)G(m_2).
\eeq
The functions $G(m_1)$ and $G(m_2)$ can be written as
\begin{align}
G(m_1)&=\int_{C_d}d^dxF_{\rm MF}(m_1^*(\bm{x})), & \int_{C_d}d^dxm_1^*(\bm{x})&=m_1, \\
G(m_2)&=\int_{C_d}d^dxF_{\rm MF}(m_2^*(\bm{x})), & \int_{C_d}d^dxm_2^*(\bm{x})&=m_2,
\end{align}
where $m_1^*(\bm{x})$ and $m_2^*(\bm{x})$ are the functions that minimize the function
$\int_{C_d}d^dxF_{\rm MF}(m(\bm{x}))$ for the fixed global variables $m_1$ and $m_2$, respectively.
Putting $\lambda x_1=y_1$ and $(1-\lambda)x_1=z_1$, we obtain
\begin{align}
&\lambda G(m_1)+(1-\lambda)G(m_2)
\nonumber \\
=&\int_0^{\lambda}dy_1\int_0^1dx_2\cdots dx_d
F_{\rm MF}\left(m_1^*\left(\frac{y_1}{\lambda},x_2,\dots,x_d\right)\right)
\nonumber \\
+&\int_0^{1-\lambda}dz_1\int_0^1dx_2\cdots dx_d
F_{\rm MF}\left(m_2^*\left(\frac{z_1}{1-\lambda},x_2,\dots,x_d\right)\right)
\nonumber \\
=&\int_0^{\lambda}dx_1\int_0^1dx_2\cdots dx_d
F_{\rm MF}\left(m_1^*\left(\frac{x_1}{\lambda},x_2,\dots,x_d\right)\right)
\nonumber \\
+&\int_{\lambda}^{1}dx_1\int_0^1dx_2\cdots dx_d
F_{\rm MF}\left(m_2^*\left(\frac{x_1-\lambda}{1-\lambda},x_2,\dots,x_d\right)\right).
\label{eq:G_convex}
\end{align}
Here, we define a function $m^*(\bm{x})$ as
\beq
m^*(\bm{x})\equiv\left\{
\begin{split}
&m_1^*\left(\frac{x_1}{\lambda},x_2,\dots,x_d\right) \qquad \text{for $0\leq x_1\leq\lambda$}, \\
&m_2^*\left(\frac{x_1-\lambda}{1-\lambda},x_2,\dots,x_d\right) \qquad \text{for $\lambda<x_1\leq 1$}.
\end{split}
\right.
\eeq
Then, the following condition is satisfied,
\beq
\int_{C_d}d^dxm^*(\bm{x})=\lambda m_1+(1-\lambda)m_2.
\eeq
Therefore, we obtain from (\ref{eq:G_convex})
\begin{align}
\lambda G(m_1)+(1-\lambda)G(m_2)
&=\int_{C_d}d^dxF_{\rm MF}(m^*(\bm{x})) \nonumber \\
&\geq G(\lambda m_1+(1-\lambda)m_2).
\end{align}
We thereby conclude that the function $G(m)$ is convex.
From Eq.~(\ref{eq:G_inequality}) and the convexity of $G(m)$,
\beq
G(m)={\rm CE}\{F_{\rm MF}(m)\}
\eeq
holds.

\bibliography{LRMF_full_ref.bib}

\begin{thebibliography}{24}
\expandafter\ifx\csname natexlab\endcsname\relax\def\natexlab#1{#1}\fi
\expandafter\ifx\csname bibnamefont\endcsname\relax
  \def\bibnamefont#1{#1}\fi
\expandafter\ifx\csname bibfnamefont\endcsname\relax
  \def\bibfnamefont#1{#1}\fi
\expandafter\ifx\csname citenamefont\endcsname\relax
  \def\citenamefont#1{#1}\fi
\expandafter\ifx\csname url\endcsname\relax
  \def\url#1{\texttt{#1}}\fi
\expandafter\ifx\csname urlprefix\endcsname\relax\def\urlprefix{URL }\fi
\providecommand{\bibinfo}[2]{#2}
\providecommand{\eprint}[2][]{\url{#2}}

\bibitem[{\citenamefont{Thirring}(1970)}]{Thirring1970}
\bibinfo{author}{\bibfnamefont{W.}~\bibnamefont{Thirring}},
  \bibinfo{journal}{Z. Phys. A} \textbf{\bibinfo{volume}{235}},
  \bibinfo{pages}{339} (\bibinfo{year}{1970}).

\bibitem[{\citenamefont{Griffiths et~al.}(1966)\citenamefont{Griffiths, Weng,
  and Langer}}]{Griffiths1966}
\bibinfo{author}{\bibfnamefont{R.~B.} \bibnamefont{Griffiths}},
  \bibinfo{author}{\bibfnamefont{C.-Y.} \bibnamefont{Weng}}, \bibnamefont{and}
  \bibinfo{author}{\bibfnamefont{J.~S.} \bibnamefont{Langer}},
  \bibinfo{journal}{Phys. Rev.} \textbf{\bibinfo{volume}{149}},
  \bibinfo{pages}{301} (\bibinfo{year}{1966}).

\bibitem[{\citenamefont{Barr\'e et~al.}(2001)\citenamefont{Barr\'e, Mukamel,
  and Ruffo}}]{Barre2001}
\bibinfo{author}{\bibfnamefont{J.}~\bibnamefont{Barr\'e}},
  \bibinfo{author}{\bibfnamefont{D.}~\bibnamefont{Mukamel}}, \bibnamefont{and}
  \bibinfo{author}{\bibfnamefont{S.}~\bibnamefont{Ruffo}},
  \bibinfo{journal}{Phys. Rev. Lett.} \textbf{\bibinfo{volume}{87}},
  \bibinfo{pages}{030601} (\bibinfo{year}{2001}).

\bibitem[{\citenamefont{Dauxois et~al.}(2002)\citenamefont{Dauxois, Ruffo,
  Arimondo, and Wilkens}}]{Lecture_notes2002}
\bibinfo{author}{\bibfnamefont{T.}~\bibnamefont{Dauxois}},
  \bibinfo{author}{\bibfnamefont{S.}~\bibnamefont{Ruffo}},
  \bibinfo{author}{\bibfnamefont{E.}~\bibnamefont{Arimondo}}, \bibnamefont{and}
  \bibinfo{author}{\bibfnamefont{M.}~\bibnamefont{Wilkens}},
  \bibinfo{journal}{Lecture Notes in Physics}  (\bibinfo{year}{2002}).

\bibitem[{\citenamefont{Dauxois et~al.}(2008)\citenamefont{Dauxois, Ruffo, and
  Cugliandolo}}]{Les_Houches_long}
\bibinfo{author}{\bibfnamefont{T.}~\bibnamefont{Dauxois}},
  \bibinfo{author}{\bibfnamefont{S.}~\bibnamefont{Ruffo}}, \bibnamefont{and}
  \bibinfo{author}{\bibfnamefont{L.}~\bibnamefont{Cugliandolo}}, in
  \emph{\bibinfo{booktitle}{Les Houches Summer School}} (\bibinfo{year}{2008}).

\bibitem[{\citenamefont{Campa et~al.}(2009)\citenamefont{Campa, Dauxois, and
  Ruffo}}]{Campa2009}
\bibinfo{author}{\bibfnamefont{A.}~\bibnamefont{Campa}},
  \bibinfo{author}{\bibfnamefont{T.}~\bibnamefont{Dauxois}}, \bibnamefont{and}
  \bibinfo{author}{\bibfnamefont{S.}~\bibnamefont{Ruffo}},
  \bibinfo{journal}{Physics Reports} \textbf{\bibinfo{volume}{480}},
  \bibinfo{pages}{57} (\bibinfo{year}{2009}).

\bibitem[{\citenamefont{Cannas et~al.}(2000)\citenamefont{Cannas, de~Magalhaes,
  and Tamarit}}]{Cannas2000}
\bibinfo{author}{\bibfnamefont{S.}~\bibnamefont{Cannas}},
  \bibinfo{author}{\bibfnamefont{A.}~\bibnamefont{de~Magalhaes}},
  \bibnamefont{and} \bibinfo{author}{\bibfnamefont{F.}~\bibnamefont{Tamarit}},
  \bibinfo{journal}{Phys. Rev. B.} \textbf{\bibinfo{volume}{61}},
  \bibinfo{pages}{11521} (\bibinfo{year}{2000}).

\bibitem[{\citenamefont{Tamarit and Anteneodo}(2000)}]{Tamarit2000}
\bibinfo{author}{\bibfnamefont{F.}~\bibnamefont{Tamarit}} \bibnamefont{and}
  \bibinfo{author}{\bibfnamefont{C.}~\bibnamefont{Anteneodo}},
  \bibinfo{journal}{Phys. Rev. Lett.} \textbf{\bibinfo{volume}{84}},
  \bibinfo{pages}{208} (\bibinfo{year}{2000}).

\bibitem[{\citenamefont{Barr{\'e}}(2002)}]{Barre2002}
\bibinfo{author}{\bibfnamefont{J.}~\bibnamefont{Barr{\'e}}},
  \bibinfo{journal}{Physica A: Statistical Mechanics and its Applications}
  \textbf{\bibinfo{volume}{305}}, \bibinfo{pages}{172} (\bibinfo{year}{2002}).

\bibitem[{\citenamefont{Barr{\'e} et~al.}(2005)\citenamefont{Barr{\'e},
  Bouchet, Dauxois, and Ruffo}}]{Barre2005}
\bibinfo{author}{\bibfnamefont{J.}~\bibnamefont{Barr{\'e}}},
  \bibinfo{author}{\bibfnamefont{F.}~\bibnamefont{Bouchet}},
  \bibinfo{author}{\bibfnamefont{T.}~\bibnamefont{Dauxois}}, \bibnamefont{and}
  \bibinfo{author}{\bibfnamefont{S.}~\bibnamefont{Ruffo}}, \bibinfo{journal}{J.
  Stat. Phys.} \textbf{\bibinfo{volume}{119}}, \bibinfo{pages}{677}
  (\bibinfo{year}{2005}).

\bibitem[{\citenamefont{Campa et~al.}(2000)\citenamefont{Campa, Giansanti, and
  Moroni}}]{Campa2000}
\bibinfo{author}{\bibfnamefont{A.}~\bibnamefont{Campa}},
  \bibinfo{author}{\bibfnamefont{A.}~\bibnamefont{Giansanti}},
  \bibnamefont{and} \bibinfo{author}{\bibfnamefont{D.}~\bibnamefont{Moroni}},
  \bibinfo{journal}{Phys. Rev. E} \textbf{\bibinfo{volume}{62}},
  \bibinfo{pages}{303} (\bibinfo{year}{2000}).

\bibitem[{\citenamefont{Campa et~al.}(2003)\citenamefont{Campa, Giansanti, and
  Moroni}}]{Campa2003}
\bibinfo{author}{\bibfnamefont{A.}~\bibnamefont{Campa}},
  \bibinfo{author}{\bibfnamefont{A.}~\bibnamefont{Giansanti}},
  \bibnamefont{and} \bibinfo{author}{\bibfnamefont{D.}~\bibnamefont{Moroni}},
  \bibinfo{journal}{J. Phys. A: Math. Theor.} \textbf{\bibinfo{volume}{36}},
  \bibinfo{pages}{6897} (\bibinfo{year}{2003}).

\bibitem[{\citenamefont{Mori}(2010)}]{Mori2010}
\bibinfo{author}{\bibfnamefont{T.}~\bibnamefont{Mori}}, \bibinfo{journal}{Phys.
  Rev. E} \textbf{\bibinfo{volume}{82}}, \bibinfo{pages}{060103}
  (\bibinfo{year}{2010}).

\bibitem[{\citenamefont{Kac et~al.}(1963)\citenamefont{Kac, Uhlenbeck, and
  Hemmer}}]{Kac1963}
\bibinfo{author}{\bibfnamefont{M.}~\bibnamefont{Kac}},
  \bibinfo{author}{\bibfnamefont{G.}~\bibnamefont{Uhlenbeck}},
  \bibnamefont{and} \bibinfo{author}{\bibfnamefont{P.}~\bibnamefont{Hemmer}},
  \bibinfo{journal}{J. Math. Phys.} \textbf{\bibinfo{volume}{4}},
  \bibinfo{pages}{216} (\bibinfo{year}{1963}).

\bibitem[{\citenamefont{Lebowitz and Penrose}(1966)}]{Lebowitz_Penrose1966}
\bibinfo{author}{\bibfnamefont{J.}~\bibnamefont{Lebowitz}} \bibnamefont{and}
  \bibinfo{author}{\bibfnamefont{O.}~\bibnamefont{Penrose}},
  \bibinfo{journal}{J. Math. Phys} \textbf{\bibinfo{volume}{7}},
  \bibinfo{pages}{98} (\bibinfo{year}{1966}).

\bibitem[{\citenamefont{Kotliar et~al.}(1983)\citenamefont{Kotliar, Anderson,
  and Stein}}]{Kotliar1983}
\bibinfo{author}{\bibfnamefont{G.}~\bibnamefont{Kotliar}},
  \bibinfo{author}{\bibfnamefont{P.~W.} \bibnamefont{Anderson}},
  \bibnamefont{and} \bibinfo{author}{\bibfnamefont{D.~L.} \bibnamefont{Stein}},
  \bibinfo{journal}{Phys. Rev. B} \textbf{\bibinfo{volume}{27}},
  \bibinfo{pages}{602} (\bibinfo{year}{1983}).

\bibitem[{\citenamefont{Katzgraber and Young}(2005)}]{Katzgraber2005}
\bibinfo{author}{\bibfnamefont{H.~G.} \bibnamefont{Katzgraber}}
  \bibnamefont{and} \bibinfo{author}{\bibfnamefont{A.~P.} \bibnamefont{Young}},
  \bibinfo{journal}{Phys. Rev. B} \textbf{\bibinfo{volume}{72}},
  \bibinfo{pages}{184416} (\bibinfo{year}{2005}).

\bibitem[{\citenamefont{Leuzzi et~al.}(2009)\citenamefont{Leuzzi, Parisi,
  Ricci-Tersenghi, and Ruiz-Lorenzo}}]{Leuzzi2009}
\bibinfo{author}{\bibfnamefont{L.}~\bibnamefont{Leuzzi}},
  \bibinfo{author}{\bibfnamefont{G.}~\bibnamefont{Parisi}},
  \bibinfo{author}{\bibfnamefont{F.}~\bibnamefont{Ricci-Tersenghi}},
  \bibnamefont{and} \bibinfo{author}{\bibfnamefont{J.~J.}
  \bibnamefont{Ruiz-Lorenzo}}, \bibinfo{journal}{Phys. Rev. Lett.}
  \textbf{\bibinfo{volume}{103}}, \bibinfo{pages}{267201}
  (\bibinfo{year}{2009}).

\bibitem[{\citenamefont{Sharma and Young}(2011)}]{Sharma2011}
\bibinfo{author}{\bibfnamefont{A.}~\bibnamefont{Sharma}} \bibnamefont{and}
  \bibinfo{author}{\bibfnamefont{A.~P.} \bibnamefont{Young}},
  \bibinfo{journal}{Phys. Rev. B} \textbf{\bibinfo{volume}{83}},
  \bibinfo{pages}{214405} (\bibinfo{year}{2011}).

\bibitem[{\citenamefont{Sherrington and Kirkpatrick}(1975)}]{Sherrington1975}
\bibinfo{author}{\bibfnamefont{D.}~\bibnamefont{Sherrington}} \bibnamefont{and}
  \bibinfo{author}{\bibfnamefont{S.}~\bibnamefont{Kirkpatrick}},
  \bibinfo{journal}{Phys. Rev. Lett.} \textbf{\bibinfo{volume}{35}},
  \bibinfo{pages}{1792} (\bibinfo{year}{1975}).

\bibitem[{\citenamefont{Binder and Young}(1986)}]{Binder_rev1986}
\bibinfo{author}{\bibfnamefont{K.}~\bibnamefont{Binder}} \bibnamefont{and}
  \bibinfo{author}{\bibfnamefont{A.~P.} \bibnamefont{Young}},
  \bibinfo{journal}{Rev. Mod. Phys.} \textbf{\bibinfo{volume}{58}},
  \bibinfo{pages}{801} (\bibinfo{year}{1986}).

\bibitem[{\citenamefont{Miyashita et~al.}(2009)\citenamefont{Miyashita,
  Rikvold, Mori, Konishi, Nishino, and Tokoro}}]{Miyashita2009}
\bibinfo{author}{\bibfnamefont{S.}~\bibnamefont{Miyashita}},
  \bibinfo{author}{\bibfnamefont{P.}~\bibnamefont{Rikvold}},
  \bibinfo{author}{\bibfnamefont{T.}~\bibnamefont{Mori}},
  \bibinfo{author}{\bibfnamefont{Y.}~\bibnamefont{Konishi}},
  \bibinfo{author}{\bibfnamefont{M.}~\bibnamefont{Nishino}}, \bibnamefont{and}
  \bibinfo{author}{\bibfnamefont{H.}~\bibnamefont{Tokoro}},
  \bibinfo{journal}{Phys. Rev. B} \textbf{\bibinfo{volume}{80}},
  \bibinfo{pages}{64414} (\bibinfo{year}{2009}).

\bibitem[{\citenamefont{Schmidt et~al.}(2001)\citenamefont{Schmidt, Kusche,
  Hippler, Donges, Kronm\"uller, von Issendorff, and Haberland}}]{Schmidt2001}
\bibinfo{author}{\bibfnamefont{M.}~\bibnamefont{Schmidt}},
  \bibinfo{author}{\bibfnamefont{R.}~\bibnamefont{Kusche}},
  \bibinfo{author}{\bibfnamefont{T.}~\bibnamefont{Hippler}},
  \bibinfo{author}{\bibfnamefont{J.}~\bibnamefont{Donges}},
  \bibinfo{author}{\bibfnamefont{W.}~\bibnamefont{Kronm\"uller}},
  \bibinfo{author}{\bibfnamefont{B.}~\bibnamefont{von Issendorff}},
  \bibnamefont{and}
  \bibinfo{author}{\bibfnamefont{H.}~\bibnamefont{Haberland}},
  \bibinfo{journal}{Phys. Rev. Lett.} \textbf{\bibinfo{volume}{86}},
  \bibinfo{pages}{1191} (\bibinfo{year}{2001}).

\bibitem[{\citenamefont{Baumann et~al.}(2010)\citenamefont{Baumann, Guerlin,
  Brennecke, and Esslinger}}]{Baumann2010}
\bibinfo{author}{\bibfnamefont{K.}~\bibnamefont{Baumann}},
  \bibinfo{author}{\bibfnamefont{C.}~\bibnamefont{Guerlin}},
  \bibinfo{author}{\bibfnamefont{F.}~\bibnamefont{Brennecke}},
  \bibnamefont{and}
  \bibinfo{author}{\bibfnamefont{T.}~\bibnamefont{Esslinger}},
  \bibinfo{journal}{Nature} \textbf{\bibinfo{volume}{464}},
  \bibinfo{pages}{1301} (\bibinfo{year}{2010}).

\end{thebibliography}

\end{document}